\documentclass[useAMS,usenatbib,usegraphicx]{mn2e}

\newcommand{\mstellar}{\ensuremath{M_{\mathrm{stellar}}}}

\newcommand{\omatter}{\ensuremath{\Omega_{\mathrm{M}}}}

\newcommand{\col}{\ensuremath{\mathcal{C}}}
\newcommand{\absm}{\ensuremath{M_B}}
\newcommand{\scriptm}{\ensuremath{\mathcal{M}_B}}
\newcommand{\absmlm}{\ensuremath{M_B^2}}
\newcommand{\absmhm}{\ensuremath{M_B^1}}
\newcommand{\deltam}{\ensuremath{\Delta m_{15}}}
\newcommand{\umega}{\ensuremath{u_M}}
\newcommand{\gmega}{\ensuremath{g_M}}
\newcommand{\rmega}{\ensuremath{r_M}}
\newcommand{\imega}{\ensuremath{i_M}}
\newcommand{\zmega}{\ensuremath{z_M}}
\newcommand{\allmegafilts}{\umega\gmega\rmega\imega\zmega}
\newcommand{\allmegasnfilts}{\gmega\rmega\imega\zmega}
\newcommand{\bmv}{\ensuremath{\left(B-V\right)}}
\newcommand{\mbcorr}{\ensuremath{m_{B}^{\mathrm{corr}}}}
\newcommand{\mB}{\ensuremath{m_{B}}}
\newcommand{\msun}{\ensuremath{M_{\sun}}}
\newcommand{\mbmodel}{\ensuremath{m_{B}^{\mathrm{mod}}}}
\newcommand{\sigint}{\ensuremath{\sigma_{\mathrm{int}}}}
\newcommand{\zcmb}{\ensuremath{z_{\mathrm{cmb}}}}

\def\aj{{AJ}}%
\def\apj{{ApJ}}%
\def\apjl{{ApJ}}%
\def\apjs{{ApJS}}%
\def\aap{{A\&A}}%
\def\aaps{{A\&AS}}%
\def\pasp{{PASP}}%
\def\pasj{{PASJ}}%
\def\nat{{Nature}}%
\def\mnras{{MNRAS}}%

\defcitealias{2006ApJ...648..868S}{S06}
\defcitealias{2009ApJ...707.1449N}{N09}
\defcitealias{2009ApJ...691..661H}{H09}

\title[SN Ia host galaxies]{The Dependence of Type Ia Supernova
  Luminosities on their Host Galaxies}
\author[M. Sullivan et al.]{M. Sullivan$^{1}$\thanks{E-mail: sullivan@astro.ox.ac.uk}, A. Conley$^{2}$, D. A. Howell$^{3,4}$, J. D. Neill$^{5}$, P. Astier$^{6}$, C. Balland$^{7,8}$,\newauthor S. Basa$^{8}$, R. G. Carlberg$^{9}$, D. Fouchez$^{10}$, J. Guy$^{6}$, D. Hardin$^{6}$, I. M. Hook$^{1,11}$,\newauthor  R. Pain$^{6}$, N. Palanque-Delabrouille$^{12}$, K. M. Perrett$^{9,13}$, C. J. Pritchet$^{14}$,\newauthor  N. Regnault$^{6}$, J. Rich$^{12}$, V. Ruhlmann-Kleider$^{12}$, S. Baumont$^{15,6}$, E. Hsiao$^{16}$,\newauthor T. Kronborg$^{6}$, C. Lidman$^{17}$, S. Perlmutter$^{18,16}$, E. S. Walker$^{19,1}$\\
  $^{1}$Department of Physics (Astrophysics), University of Oxford, DWB, Keble Road, Oxford OX1 3RH, UK\\
  $^{2}$Department of Astrophysical and Planetary Sciences, University of Colorado, Boulder, CO 80309-0391, USA\\
  $^{3}$Las Cumbres Observatory Global Telescope Network, 6740 Cortona Dr.,
  Suite 102, Goleta, CA 93117, USA\\
  $^{4}$Department of Physics, University of California, Santa Barbara, Broida
  Hall, Mail Code 9530, Santa Barbara, CA 93106-9530, USA\\
  $^{5}$California Institute of Technology, 1200 E. California Blvd., Pasadena, CA 91125, USA\\
  $^{6}$LPNHE, Universit\'e Pierre et Marie Curie Paris 6,
  Universit\'e Paris Diderot Paris 7, CNRS-IN2P3,
  4 place Jussieu, 75252 Paris Cedex 05, France.\\
  $^{7}$University Paris 11, Orsay, F-91405, France\\
  $^{8}$LAM, Pole de l'Etoile Site de Chateau-Gombert, 38 rue Frederic Joliot-Curie, 13388 Marseille Cedex 13, France\\
  $^{9}$Department of Astronomy and Astrophysics, University of Toronto, 50 St. George Street, Toronto ON M5S 3H4, Canada\\
  $^{10}$CPPM, Aix-Marseille Universit\'e, CNRS/IN2P3, 13288 Marseille Cedex 9, France\\
  $^{11}$ INAF - Osservatorio Astronomico di Roma, via Frascati 33, 00040 Monteporzio (RM), Italy\\
  $^{12}$CEA/Saclay, DSM/Irfu/Spp, 91191 Gif-sur-Yvette Cedex, France\\
  $^{13}$ Network Information Operations, DRDC Ottawa, 3701 Carling Avenue, Ottawa, ON, K1A 0Z4, Canada\\
  $^{14}$Department of Physics and Astronomy, University of Victoria, P.O. Box 3055 STN CSC, Victoria BC V8T 1M8, Canada\\
  $^{15}$LPSC, CNRS-IN2P3, 53 rue des Martyrs, 38026 Grenoble Cedex, France\\
  $^{16}$Lawrence Berkeley National Laboratory, Mail Stop 50-232, Lawrence Berkeley National Laboratory, 1 Cyclotron Road, Berkeley, CA 94720, USA\\
  $^{17}$Anglo-Australian Observatory, P.O. Box 296, Epping, NSW 1710, Australia\\
  $^{18}$Department of Physics, University of California, 366 LeConte Hall MC 7300, Berkeley, CA 94720-7300, USA\\
  $^{19}$Scuola Normale Superiore, Piazza dei Cavalieri 7, 56126 Pisa, Italy\\
}
\begin{document}

\pagerange{\pageref{firstpage}--\pageref{lastpage}}

\maketitle

\label{firstpage}

\clearpage

\begin{abstract}

  Precision cosmology with Type Ia supernovae (SNe Ia) makes use of
  the fact that SN Ia luminosities depend on their light-curve shapes
  and colours.  Using Supernova Legacy Survey (SNLS) and other data,
  we show that there is an additional dependence on the global
  characteristics of their host galaxies: events of the same
  light-curve shape and colour are, on average, 0.08\,mag
  ($\simeq4.0\sigma$) brighter in massive host galaxies (presumably
  metal-rich) and galaxies with low specific star-formation rates
  (sSFR).  These trends do not depend on any assumed cosmological
  model, and are independent of the SN light-curve width: both fast
  and slow-declining events show the same trends.  SNe Ia in galaxies
  with a low sSFR also have a smaller slope (``$\beta$'') between
  their luminosities and colours with $\sim$2.7$\sigma$ significance,
  and a smaller scatter on SN Ia Hubble diagrams (at 95\% confidence),
  though the significance of these effects is dependent on the reddest
  SNe.  SN Ia colours are similar between low-mass and high-mass
  hosts, leading us to interpret their luminosity differences as an
  intrinsic property of the SNe and not of some external factor such
  as dust. If the host stellar mass is interpreted as a metallicity
  indicator using galaxy mass--metallicity relations, the luminosity
  trends are in qualitative agreement with theoretical predictions. We
  show that the average stellar mass, and therefore the average
  metallicity, of our SN Ia host galaxies decreases with redshift.
  The SN Ia luminosity differences consequently introduce a systematic
  error in cosmological analyses, comparable to the current
  statistical uncertainties on parameters such as $w$, the equation of
  state of dark energy. We show that the use of two SN Ia absolute
  magnitudes, one for events in high-mass (metal-rich) galaxies, and
  one for events in low-mass (metal-poor) galaxies, adequately
  corrects for the differences.  Cosmological fits incorporating these
  terms give a significant reduction in $\chi^2$ (3.8--4.5$\sigma$);
  linear corrections based on host parameters do not perform as well.
  We conclude that all future SN Ia cosmological analyses should use a
  correction of this (or similar) form to control demographic shifts
  in the underlying galaxy population.

\end{abstract}

\begin{keywords}
supernovae: general -- cosmology: observations -- distance scale.
\end{keywords}

\section{Introduction}

As calibrateable standard candles, Type Ia supernovae (SNe Ia) provide
a direct route to understanding the nature of the dark energy that
drives the accelerated expansion of the Universe. Yet, the
relationships that allow the calibration of their peak luminosities,
and hence permit their cosmological use, remain purely empirical.
Relations between the width of the SN Ia light curve and peak
luminosity \citep{1993ApJ...413L.105P} and between the SN Ia optical
colours and luminosity
\citep[e.g.][]{1996ApJ...473...88R,1998A&A...331..815T} reduce the
scatter in their peak magnitudes to $\sim$0.15\,mag
\citep{2007ApJ...659..122J,2007A&A...466...11G,2008ApJ...681..482C}.
As the available SN Ia samples increase in both size and quality, and
the dark energy constraints they provide become correspondingly more
statistically precise, it is increasingly important that the validity
of these calibrating relationships is robustly examined.

The observed properties of SNe Ia are known to correlate with the
physical parameters defining their host galaxy stellar populations.
SNe Ia are more than an order of magnitude more common (per unit
stellar mass) in actively star-forming or morphologically late-type
galaxies than in passive or elliptical systems
\citep{2005A&A...433..807M,2006ApJ...648..868S}.  SNe Ia in elliptical
or passively evolving systems are also intrinsically fainter, with
narrower, faster (or lower ``stretch''), light curves
\citep{1995AJ....109....1H,1996AJ....112.2398H,1999AJ....117..707R,2000AJ....120.1479H,2006ApJ...648..868S}.
Though this effect is corrected for by the light-curve shape
correction, the amount of star formation activity in the universe
increases with redshift, and these differences lead to an observed
``demographic shift'' in mean SN Ia properties.  A greater fraction of
intrinsically luminous, wider light-curve events in the distant
universe are seen compared to that observed locally
\citep{2007ApJ...667L..37H}. These photometric differences are also
partially reflected in SN Ia spectra, with SNe Ia in spiral galaxies
showing weaker intermediate mass element line strengths than those in
elliptical galaxies \citep{2008A&A...477..717B,2009A&A...507...85B},
and a corresponding evolution in the mean SN Ia spectrum with redshift
\citep{2009ApJ...693L..76S}.

There are suggestions that these effects may be the result of multiple
astrophysical channels capable of producing SN Ia explosions
\citep[e.g.][]{2005ApJ...629L..85S,2006MNRAS.370..773M}.  In
particular, delay-time distributions with distinct ``prompt'' and
``delayed'' components, or with a wide range of delay-times, match
most observational datasets well
\citep{2006MNRAS.370..773M,2006ApJ...648..868S,2008ApJ...683L..25P,2008PASJ...60.1327T},
though the minimum age for the prompt systems remains controversial
\citep{2008A&A...492..631A,2009ApJ...707...74R} with some evidence
that ``prompt'' SNe Ia occur more frequently in metal-poor systems
\citep{2009ApJ...704..687C}. The use of SNe Ia as precision
cosmological probes therefore depends on establishing that the
demographic shifts, or existence of multiple channels to a SN Ia, do
not impact on the light-curve-width/colour/luminosity relationships.
If these relations show environmental dependence, then the task of
calibrating SNe Ia for cosmology becomes substantially more
challenging \citep[e.g.][]{2008ApJ...684L..13S,2009arXiv0912.0929K}.

A second complication arises from the (poorly understood) colour
corrections applied to SN Ia luminosities. Redder SNe Ia appear
fainter than their bluer counterparts, but the slope of the
relationship between SN Ia colour and magnitude is inconsistent with
the ratio of total-to-selective absorption appropriate for the diffuse
interstellar medium of the Milky Way ($R_V=3.1$).  Multiple studies of
different SN Ia samples indicate that the effective $R_V$ inferred
from normal SNe Ia is smaller than 3.1
\citep[e.g.,][]{1998A&A...331..815T,2006A&A...447...31A,2006AJ....131.1639K},
and the \textit{assumption} of $R_V=3.1$, even after light curve shape
corrections, leads to serious systematic error problems such as a
spurious ``Hubble Bubble''
\citep{2007ApJ...659..122J,2007ApJ...664L..13C}. The reason for this
low effective $R_V$ is not well understood. Although uncorrected
intrinsic variations in the SN Ia population could play a role
\citep[e.g.][]{2009Natur.460..869K}, some dust extinction must also
affect the SN Ia luminosities and colours, and this may vary by
environment. Furthermore, the exact value of $R_V$ obtained is
sensitive to method used to determine it, with lower $R_V$ obtained
when fitting linear relations between SN Ia luminosities, colours, and
light curve widths \citep{2010AJ....139..120F}, perhaps due to
intrinsic variation in SN Ia colours that correlates with luminosity
but not light-curve width. Current knowledge of SNe Ia is not
sufficient to separate and correct for both intrinsic
colour--luminosity and dust-induced colour--luminosity effects in
cosmological SN Ia samples.

Examining how SN Ia luminosities vary with environment after light
curve shape and colour corrections can place constraints on the degree
of these possible variations.  Early studies showed little evidence
that corrected SN Ia luminosities varied with host galaxy morphologies
\citep[e.g.,][]{1999ApJ...517..565P,1999AJ....117..707R,2003MNRAS.340.1057S,2003AJ....126.2608W,2005ApJ...634..210G},
though these tests used relatively small samples of events ($\la50$),
in some cases from the first-generation of SN Ia cosmological samples
before dense multi-colour light curves were routinely obtained.

More recent analyses, using larger, well-observed samples, have shown
tentative evidence for variation.  \citet{2009ApJ...700.1097H} found
$\simeq2\sigma$ evidence that SNe Ia in morphologically E/S0 galaxies
are brighter than those in later-type spirals after light-curve shape
and colour corrections. Extending beyond simple host galaxy
morphologies to more physically motivated variables gives further
tantalising suggestions of variation.  \citet{2008ApJ...685..752G}
found evidence for a correlation between Hubble diagram residual and
host galaxy stellar metallicity in a sample of 17 local SNe Ia located
in E/S0 galaxies, in the sense that fainter SNe Ia after correction
were found in metal poor systems (note this is the reverse of the
originally published trend due to an error in the original analysis;
P. Garnavich, private communication).  \citet{2009ApJ...691..661H}
used 55 SNe Ia from the first year of the SNLS and showed no
significant correlation between Hubble residual and host galaxy
metallicity, albeit using host gas-phase metallicities inferred from
average galaxy stellar-mass--metallicity relations, a less direct
measure of metallicity.  \citet{2009arXiv0912.0929K} have shown a
relation between host galaxy stellar mass and Hubble residual, in the
sense that more massive systems host brighter SNe Ia after light curve
shape and colour corrections.  Under the assumption that more massive
galaxies are metal rich, this trend is consistent with the revised
\citet{2008ApJ...685..752G} result.

In this paper, we use a sample of 282 high redshift SNe Ia discovered
and photometrically monitored by the Canada-France-Hawaii Telescope
(CFHT) as part of the Supernova Legacy Survey (SNLS), and which form
the SNLS ``three-year'' sample (SNLS3). Using deep optical imaging of
their host galaxies taken over the duration of the survey, we place
constraints on their recent star-formation activity, stellar masses
(and hence inferred metallicity), and compare to the photometric
properties of the SNe Ia that they host. In particular, we search for
evidence that the corrected SN Ia luminosities correlate with these
host properties, indicating possible systematic errors in the light
curve fitting framework that underpins their cosmological use.  We
compare with the properties of a sample of lower-redshift SNe Ia taken
from the literature.

A plan of the paper follows. In $\S$~\ref{sec:data} we introduce the
SN Ia sample and the data available on their host galaxies.
$\S$~\ref{sec:sn-properties-as} investigates how the SN Ia light curve
widths and colours of these SNe Ia varies according to their host
galaxy properties, and in $\S$~\ref{sec:lum-depend-trends} we compare
their corrected luminosities to the host properties. We discuss the
results, including the cosmological implications, in
$\S$~\ref{sec:discussion}, and conclude in $\S$~\ref{sec:conclusions}.
Throughout, where relevant we assume a flat $\Lambda$CDM cosmological
model with $\omatter=0.256$ (the reason for this non-standard choice
is explained in $\S$~\ref{sec:lum-depend-trends}) and
$\mathrm{H}_0$=70\,km\,s$^{-1}$\,Mpc$^{-1}$ assumed in all quoted
absolute magnitudes.  All magnitudes are given on the
\citet{1992AJ....104..340L} photometric system as described in
\citet{2009A&A...506..999R}.

\section{Type Ia Supernova and host galaxy data}
\label{sec:data}

We begin by introducing the SN Ia samples that we will use in this
paper, and the associated data available for their host galaxies.

\subsection{The SN Ia samples}
\label{sec:snsample}

The high-redshift SN Ia data is taken from the Supernova
Legacy Survey (SNLS). This used optical imaging data taken as part of
the deep component of the five-year Canada-France-Hawaii Telescope
Legacy Survey (CFHT-LS) using the square-degree ``MegaCam'' camera
\citep{2003SPIE.4841...72B}, located in the prime focus environment
``MegaPrime'' on the CFHT. The ``deep'' component of CFHT-LS conducted
repeat imaging of 4 low Galactic-extinction fields, time-sequenced
with observations conducted every 3--4 nights in dark and grey time.
Four filters, \allmegasnfilts, were used allowing the construction of
high-quality multi-colour SN light curves; \umega\ data were also
taken but are not time-sequenced.  On each night of observations, the
data were searched using two independent pipelines, and an amalgamated
candidate list produced\footnote{Candidates can be found at
  \texttt{http://legacy.astro.utoronto.ca/}}
\citep[see][]{2010perrettrta}. Spectroscopic follow-up confirmed SN
types and measured redshifts to allow SNe Ia to be placed on a Hubble
diagram.

In this paper we use SNe Ia belonging to the three-year sample, SNLS3;
this includes all SNe Ia discovered up until the end of July 2006. The
light curves and other details for these SNe can be found in
\citet{2009guylcs}, and spectroscopic information is taken from
\citet{2005ApJ...634.1190H}, \citet{2008A&A...477..717B},
\citet{2008ApJ...674...51E}, \citet{2009A&A...507...85B}, and Walker
et al. (2010). The full SNLS3 sample consists of 282 SNe Ia, however
we exclude some of these events from some of our analyses:
\begin{enumerate}
\item Only SNLS SNe~Ia with an identified
  ($\S$~\ref{sec:sn-host-galaxy}) host galaxy (272 events) are
  considered \citep[for a discussion of the identification procedures,
  see][hereafter S06]{2006ApJ...648..868S},
\item Only SNe Ia passing light curve quality cuts
  \citep[e.g.][]{2008ApJ...681..482C,2009guylcs} are used -- there
  must be sufficient photometric coverage to derive reliable peak
  luminosities, light curve widths, and colours \citep[see details
  in][]{2009guylcs},
\item We only consider normal SNe Ia with light-curve parameters in
  the range considered for cosmological fits -- our motivation in this
  paper is to assess the cosmological impact of any host galaxy
  dependent trends. Specifically, we require that the stretch of the
  SN be greater than 0.80, and that the colour be less than 0.30 (see
  $\S$~\ref{sec:sn-light-curve} for a discussion of the meaning of
  these parameters). We also discard $>3\sigma$ outliers from the best
  fitting cosmological model. 231 events pass the light-curve coverage
  and SN parameter cuts.
\item In analyses where we search for trends in the SNLS data, we use
  only SNe~Ia at redshift $z\leq0.85$ (195 events). At these lower
  redshifts, both the SN and host galaxy photometry are higher
  signal-to-noise and their photometric parameters better measured.
  The SNLS Malmquist biases are also smaller \citep{2010perrettrta}.
\end{enumerate}

Where relevant, we also use samples of SNe Ia from the literature. We
construct a sample of low-redshift SNe Ia from the compilation of
\citet{2009conleysys}, which includes SNe Ia from a variety of sources
\citep[primarily][]{1996AJ....112.2408H,1999AJ....117..707R,2006AJ....131..527J,2009ApJ...700..331H,2010AJ....139..519C}.
We apply bulk-flow peculiar velocity corrections to the SN magnitudes
and redshifts, placing the redshifts in the CMB-frame (\zcmb)
following \citet{2007ApJ...661L.123N}, but with updated models
\citep{2009conleysys}. The accuracy of these corrections is estimated
to $\pm$150 km s$^{-1}$, which we propagate into the SN magnitude
errors in cosmological fits. We only use SNe Ia in the smooth Hubble
flow, here defined as $\zcmb\geq0.01$, and apply the same light curve
quality cuts as for the SNLS sample. There are 110 low-redshift objects in
total. We also use the \textit{HST}-discovered sample of
\citet{2004ApJ...607..665R} and \citet{2007ApJ...659...98R}, hereafter
the R07 sample. We select 24 SNe Ia at $z>0.9$ from this sample to
increase our redshift lever arm above $z=1$.

\subsection{SN Light curve fitting}
\label{sec:sn-light-curve}

In its current application, SN Ia cosmology depends on two corrections
to ``raw'' SN Ia peak luminosities that when applied reduce the
dispersion in their peak magnitudes. The first is the
light-curve-shape/luminosity relation
\citep[e.g.][]{1993ApJ...413L.105P}: brighter SNe Ia tend to have
wider, longer-duration light curves (higher stretch) than their
fainter counterparts. The second is a colour correction: brighter SNe
Ia tend to have bluer colours, whilst fainter SNe tend to be redder
\citep{1996ApJ...473...88R,1998A&A...331..815T}.  Together the
application of these corrections can yield distance estimates precise
to $\simeq6$ per cent. These corrections are applied in different ways
depending on whether a technique is a distance estimator \citep[e.g.,
MLCS2k2;][]{2007ApJ...659..122J} or a light curve fitter \citep[e.g.
SALT or SiFTO;][]{2007A&A...466...11G,2008ApJ...681..482C}, though the
underlying principle is the same in both approaches.

In this paper, we primarily use the SiFTO light curve fitter
\citep{2008ApJ...681..482C} and compare our results to SALT2
\citep{2007A&A...466...11G} where appropriate. In general SALT2 and
SiFTO give very similar results when trained on the same SN sample --
a full discussion can be found in \citet{2009guylcs}. Both fitters
have been retrained and improved since the original published versions
using SNLS and low-redshift data. The product of both fitters for each SN
is a rest-frame $B$-band apparent magnitude (\mB), a stretch ($s$)
measurement, and a colour estimate (\col), together with associated
errors and covariances (SALT2 uses the broadly equivalent $x_1$
parameter in place of $s$).  Throughout, the SN~Ia colour, \col, is
defined as the rest-frame \bmv\ colour of the SN at the time of
maximum light in the rest-frame $B$-band. We refit all SNe Ia using
these light curve fitters to ensure that the different samples can be
placed on the same distance scale. A full discussion of their
application to the current dataset, together with their tabulated
output, can be found in \citet{2009guylcs} and \citet{2009conleysys}.

\subsection{SN host galaxy photometry}
\label{sec:sn-host-galaxy}

Our SNLS host galaxy photometry comes in the optical from the CFHT-LS
(\allmegafilts) and in the near infra red (IR) from the WIRcam Deep
Survey (WIRDS; Bielby et al. in prep.) of a sub-section of the CFHT-LS
fields ($J$, $H$, $K_s$). The identification procedure for the SNLS SN
Ia host galaxies is the same as that in
\citetalias{2006ApJ...648..868S}.  Photometry is performed by
\texttt{SExtractor} \citep{1996A&AS..117..393B} using FLUX\_AUTO
photometry running in dual-image mode, detecting from the deep \imega\
stack and measuring from each of the optical and near-IR filters in
turn (60\% of our SN Ia hosts have near-IR data).  Each stack has a
similar image quality and hence the same physical aperture is used in
each filter.  In about $\simeq$3\% of cases no host galaxy can be
identified. This could be because the SN lies far from the centre of
its host galaxy leading to ambiguity in the correct choice of host, or
because the host lies below the CFHT-LS flux limits. We discard these
objects.  Weight maps are used to determine the measurement errors,
and in the optical, the photometric zeropoints are generated using a
comparison to the tertiary standard star lists of
\citet{2009A&A...506..999R}.  No SN light is present in the optical
stacks which are constructed on a per season basis
\citepalias{2006ApJ...648..868S}.

For the low-redshift sample, we use host galaxy photometry recently
compiled by \citet[][hereafter N09]{2009ApJ...707.1449N}, including
ultraviolet data from the GALEX (Galaxy Evolution Explorer) satellite,
and optical photometry from the third reference catalog of bright
galaxies \citep[RC3;][]{1994AJ....108.2128C} and the Sloan Digital Sky
Survey (SDSS). The photometry for this sample was carefully performed
in matched apertures with foreground contaminating stars masked.
Though these data span a different observed wavelength range compared
to the SNLS host photometry, in the rest-frame the wavelength range
covered is reasonably similar: the maximal range is 1500--9000\AA\ for
the low-redshift sample (though frequently the available data span a
smaller range than this), and 2400--13000\AA\ for the SNLS sample (at
$z=0.6$).  Note that not all the low-redshift SN hosts have publicly
available host photometry: only 69 (of 110; 63\%) low-redshift SNe Ia
have sufficient data and are carried forward in the analysis.  The
missing low-redshift SNe Ia are due to incomplete GALEX and SDSS
coverage, rather than the host galaxies being too faint to be detected
by the two surveys.

For the R07 sample, we use photometry taken from the Great
Observatories Origins Deep Survey (GOODS) \textit{HST} data
\citep{2004ApJ...600L..93G}, taken with the Advanced Camera for
Surveys (ACS). Data are available in four filters: F435W (broadly
equivalent to a $B$ filter), F606W ($V$), F814W ($i'$) and F850LP
($z'$). All of the R07 SNe have ACS coverage, and again
\texttt{SExtractor} FLUX\_AUTO photometry is used. We also use $J$,
$H$ and $K$ imaging taken as part of the GOODS NICMOS survey
\citep[e.g.][]{2008ApJ...687L..61B}, as well as ground-based data
\citep{2010A&A...511A..50R}.

\subsection{Host galaxy parameter estimation}
\label{sec:galaxy-param-estim}

The method to estimate physical parameters of the SN Ia host galaxies
is similar to that in \citetalias{2006ApJ...648..868S} which used the
photometric redshift code Z-PEG \citep{2002A&A...386..446L} based upon
the P\'EGASE.2 spectral synthesis code
\citep[e.g.,][]{1997A&A...326..950F}. We use an expanded set of 15
exponentially declining star formation histories (SFHs) with SFR
$\propto \mathrm{exp}^{-t/\tau}$, where $t$ is time and $\tau$ is the
e-folding time, each with 125 age steps.  The internal P\'EGASE.2 dust
prescriptions are not used, and instead a foreground dust screen
varying from $E(B-V)=0$ to 0.30 in steps of 0.05 is added.  With the 7
different foreground dust screens, this gives a total of 105 unique
evolving spectral energy distributions (SEDs). The metallicity of the
stellar population evolves consistently following the standard
P\'EGASE.2 model with an initial value of 0.0004, and the standard
P\'EGASE.2 nebular emission prescription is added.

Z-PEG is used to locate the best-fitting SED model (in a $\chi^2$
sense), with the redshift fixed at the CMB-frame redshift of the SN.
Only solutions younger than the age of the Universe at each SN
redshift are permitted. The current stellar mass in stars (\mstellar,
measured in \msun), the recent star formation rate (SFR, in $\msun
\mathrm{yr}^{-1}$, averaged over the last 0.25Gyr before the best
fitting time step), and the specific star formation rate \citep[sSFR,
the SFR per unit stellar mass with units of $\mathrm{yr}^{-1}$,
e.g.][]{1997ApJ...489..559G} are all recorded. Error bars on these
parameters are taken from their range in the set of solutions that
have a similar $\chi^2$ \citepalias[as in][]{2006ApJ...648..868S}.
Note that we do not measure the instantaneous SFR as we only fit
broad-band photometry. We refit all the
\citetalias{2009ApJ...707.1449N} photometry to ensure the exact same
library SEDs are used for all hosts.

We use a \citet{1992A&A...265..499R} initial mass function (IMF), the
P\'EGASE.2 default, throughout. Our results are not sensitive to this
choice -- we have repeated our analysis in full with both the more
standard \citet{1955ApJ...121..161S} IMF, and with a
\citet{2003ApJ...593..258B} IMF, and find similar results, though the
\mstellar\ and SFRs derived for the host galaxies have (expected)
small mean offsets when using different IMFs. In detail, the use of a
Salpeter IMF gives systematically larger host masses by 0.04 dex
(smaller masses by 0.16 dex for the B\&G IMF), and smaller SFRs by
0.04 dex (smaller by 0.16 dex for B\&G), with scatter of around 0.1dex
in each comparison. These differences are not a function of \mstellar,
SFR or redshift and so have a negligible impact on our conclusions.

As only $\simeq$60\% of our SNLS SN Ia hosts have observer-frame
near-IR data, we compare the derived properties with and without these
data to check for potential biases in the remaining 40\% of objects.
The mean difference in \mstellar\ (defined as
$\mstellar^{\mathrm{OPT}}$-$\mstellar^{\mathrm{IR}}$) is 0.001 dex
(r.m.s. 0.15) and in the recent SFR the mean difference
(SFR$^{\mathrm{OPT}}$-SFR$^{\mathrm{IR}}$) is $-0.18$ dex (r.m.s.
0.44), in the sense that excluding the IR data leads to smaller SFRs.
Thus we find no evidence that the near-IR data leads to systematically
different \mstellar, and mild evidence that including these data leads
to larger SFRs. The differences in SFR do not follow a Gaussian
distribution; instead the difference is centred around zero but with a
long tail to negative differences; therefore we choose not to apply
the mean offset to the 40\% of hosts with no IR data. There is no
evidence for any redshift dependent trend. Information on the derived
properties for the SNLS, low-redshift and R07 hosts can be found in
Table~\ref{tab:hostprops}.

\begin{table*}
\centering
\caption{Properties for the SNLS, low-redshift and R07 host galaxies. The full table can be found in the electronic version of the journal.\label{tab:hostprops}}
\begin{tabular}{ccccc}
\hline
SN Name & \zcmb & \imega\ & LOG \mstellar\ & LOG SFR\\
& & & (\msun) & ($\msun \mathrm{yr}^{-1}$)\\
\hline
03D1ar & 0.408 &    19.57 & $10.37^{+0.28}_{-0.10}$ &  $1.55^{+0.27}_{-0.77}$ \\
03D1au & 0.504 &    21.87 &  $9.55^{+0.13}_{-0.09}$ &  $0.63^{+0.58}_{-0.63}$ \\
03D1aw & 0.582 &    22.53 &  $9.21^{+0.04}_{-0.14}$ &  $0.87^{+0.03}_{-0.37}$ \\
03D1ax & 0.496 &    18.49 & $11.61^{+0.15}_{-0.08}$ &                $<-3.00$ \\
03D1bk & 0.865 &    20.21 & $11.47^{+0.08}_{-0.03}$ &  $0.46^{+0.04}_{-3.46}$ \\
03D1bp & 0.347 &    18.72 & $10.85^{+0.20}_{-0.05}$ &  $0.71^{+0.33}_{-0.33}$ \\
03D1co & 0.679 &    23.94 &  $8.69^{+0.50}_{-0.06}$ & $-0.06^{+0.66}_{-0.43}$ \\
03D1dt & 0.612 &    21.91 &  $9.76^{+0.08}_{-0.11}$ &  $0.41^{+0.16}_{-0.16}$ \\
03D1ew & 0.868 &    26.37 &  $8.58^{+0.70}_{-0.91}$ & $-0.96^{+1.32}_{-2.04}$ \\
03D1fc & 0.332 &    19.55 & $10.41^{+0.02}_{-0.05}$ &  $0.47^{+0.25}_{-0.24}$ \\

\hline
\end{tabular}
\end{table*}

\begin{figure*}
\centering
\includegraphics[width=0.5824\textwidth]{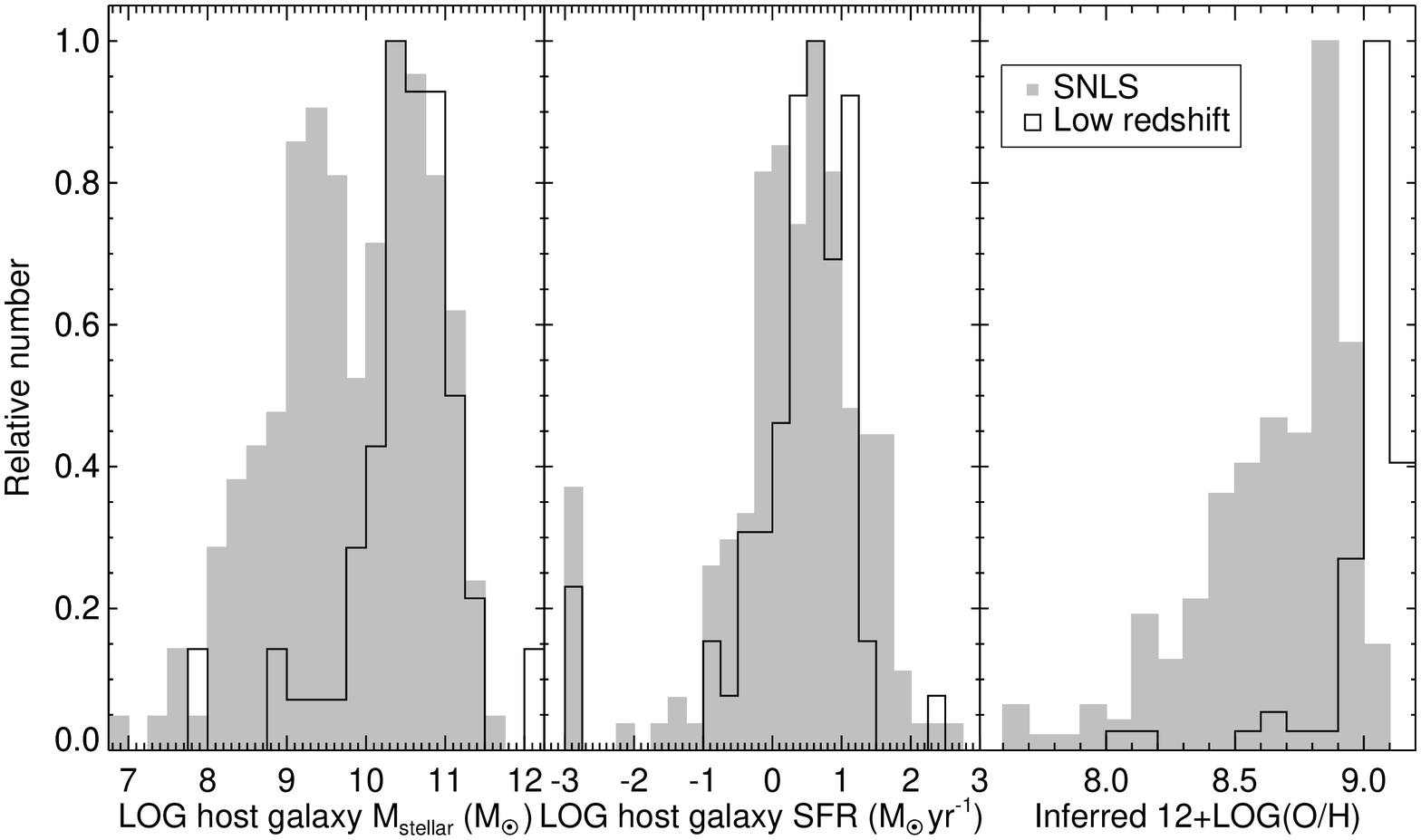}
\includegraphics[width=0.4076\textwidth]{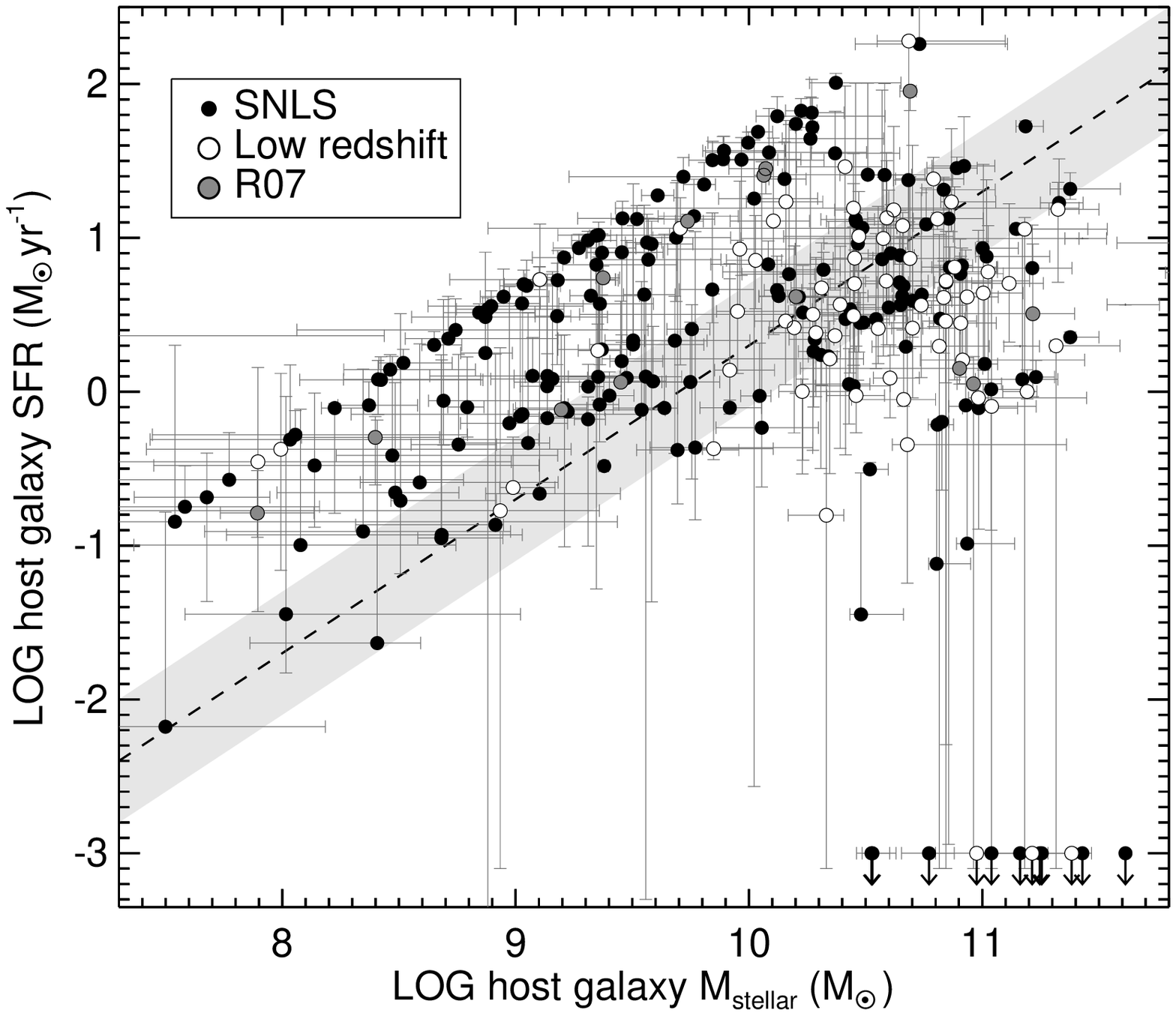}
\caption{The distribution of SN~Ia host galaxies in \mstellar, SFR and
  inferred gas-phase metallicity.  Left: Histograms in \mstellar, SFR,
  and metallicity; low-redshift (open histogram) versus SNLS (filled
  histogram). Right: The distribution in the \mstellar--SFR plane. The
  dashed line is a line of constant sSFR used as the default split to
  separate high sSFR and low sSFR galaxies.  The light-grey shaded
  areas show the range over which the default splitting value is
  varied in Section~\ref{sec:test-univ-nuis}. SNLS SN Ia hosts are
  shown as filled circles, low-redshift hosts as white circles, R07
  hosts as grey circles. Note that the apparent diagonal ridgeline in
  the \mstellar--SFR plane is a result of the maximum sSFR being
  reached in the model SFHs, a disadvantage of the simple SFHs
  considered here.\label{fig:sfrmass}}
\end{figure*}

The \mstellar\ and SFR distributions for the SNLS and low-redshift
samples are shown in Fig.~\ref{fig:sfrmass}, together with the
distribution of galaxies in the SFR--\mstellar\ plane.  As might be
expected, galaxies with the smallest sSFRs tend to be the most massive
systems, with the lowest mass systems universally consistent with
strong star formation activity. As previously highlighted by
\citetalias{2009ApJ...707.1449N}, the SNLS and low-redshift hosts show
quite different distributions in \mstellar\ (and to a lesser extent
SFR): The low-redshift SNe are drawn from more massive host galaxies.
This is almost certainly due to selection biases.  SNLS is a rolling
search which will locate SNe Ia in any type of host galaxy in which
they explode, and, modulo any small spectroscopic follow-up bias, this
range will be reflected in the cosmological sample. At low-redshift,
most SNe Ia are drawn from galaxy-targeted searches which search known
(and typically bright/massive) galaxies, consequently the most massive
systems will be over-represented.

Following \citet{2009ApJ...691..661H}, we convert the \mstellar\ mass
estimates into metallicities using average \mstellar--metallicity
(\mstellar--$Z$) relations. As the universe ages, galaxies will become
more massive via merging processes, and more metal rich following
chemical enrichment and decreased metal loss. We use a relation
between gas-phase metallicity, explicitly the nebular oxygen abundance
relative to hydrogen, $\mathrm{O}/\mathrm{H}$, and \mstellar\ derived
from SDSS galaxies \citep{2004ApJ...613..898T}. We use units of
$12+\log(\mathrm{O}/\mathrm{H})$, where the solar abundance is
$\simeq$8.69. This \mstellar--$Z$ relation is observed to evolve with
redshift, being shifted towards high \mstellar\ or lower metallicities
at higher redshift
\citep[e.g.][]{2005ApJ...635..260S,2009A&A...495...53L}.  To account
for this effect, we use the evolution measured by
\citet{2009A&A...495...53L} relative to the
\citet{2004ApJ...613..898T} relation. (Note that the use of a
\mstellar--$Z$ relation that evolves with redshift will exacerbate the
difference in \mstellar\ between low redshift and SNLS hosts when
expressed as a metallicity.) Though the calibration of nebular-line
metallicity diagnostics is a complex topic
\citep[e.g.][]{2008ApJ...681.1183K}, the \textit{exact} calibration
scale does not concern us here, so long as we can place all our hosts
on the same relative system to search for variations in the SN sample.

The \citet{2004ApJ...613..898T} relation is derived for gas-phase
metallicity, and may not be applicable in elliptical galaxies with
little cold gas where a stellar metallicity may be more appropriate.
In principle, we could also use a stellar metallicity--mass relation
\citep[e.g.,][]{2005MNRAS.362...41G}. However, this relation has only
been measured at low-redshift, and any (expected) evolution with
redshift has not been constrained observationally -- stellar
metallicities are significantly more difficult to measure than
gas-phase metallicities, requiring absorption line measures instead of
emission lines. Therefore we restrict ourselves to the gas-phase
metallicity, but note that as the mechanism for retaining metals is
the depth of the galaxy gravitational potential well, ellipticals (or
passively evolving galaxies) should follow the same general trend
between stellar metallicity and \mstellar\
\citep{2005MNRAS.362...41G}, with stellar metallicity and gas-phase
nebular metallicity well correlated
\citep[e.g.][]{2005MNRAS.358..363C,2005MNRAS.362...41G}.

The broad-band SED fitting approach used here is a relatively crude
way to determine galaxy properties. However, our choice is limited as,
for the SNLS sample at least, we only have optical \allmegafilts\
magnitudes (albeit measured from extremely deep and well-calibrated
images) and some near-IR information for the hosts. In particular,
spectroscopic diagnostics used in many other galaxy analyses are not
available. Though each SN Ia does have a spectrum containing some host
galaxy spectral information, these form a very heterogeneous set
invariably contaminated by SN light, making the accurate measurement
of host spectral features impossible. We did explore adding GALEX
ultraviolet data to our SNLS host fits \citepalias[as
in][]{2009ApJ...707.1449N}; however the relatively poor resolution of
the GALEX imaging led to substantial source confusion for most of our
galaxies. We also experimented with the use of more complex
(``stochastic'') SFHs involving bursts of star formation superimposed
on smooth underlying SFHs as in other studies of SN Ia or gamma-ray
burst host galaxies
\citep[e.g.][]{2009ApJ...691..182S,2009MNRAS.397..717S}; however the
increased number of free parameters (e.g., burst strength, burst
duration, burst age, etc.) makes the problem quite degenerate. Hence,
our host galaxy parameters are only representative of the dominant
star formation episode in each host. We note that the use of
stochastic SFHs results in increased derived stellar masses of
galaxies by about 0.14\,dex
\citep{2007A&A...474..443P,2009A&A...495...53L}.

Where required in our analysis to examine the dependence of SN Ia
properties on environmental properties, we classify the SN Ia host
galaxies into different groups.  The first split is performed
according to their sSFR: galaxies with $\log(\mathrm{sSFR})<-9.7$ with
smaller amounts of star formation relative to their stellar masses are
classified as low-sSFR, and those with a larger sSFR are classified as
high-sSFR. This split will separate those galaxies dominated by young
stellar populations from those comprised of older stellar populations.
The second split is based upon the host galaxy \mstellar\
($\log(\mstellar)>10.0$ are defined as high mass), with an extension
to the inferred gas-phase metallicity ($Z>8.8$ are defined as metal
rich).  (We consider the effect of varying these split points in later
sections.) Approximately 30 per cent of SNe Ia in our SNLS sample are
found in low-sSFR systems according to the definition above.

\section{SN~Ia properties as a function of host galaxy properties}
\label{sec:sn-properties-as}

In this paper we correct SN Ia peak magnitudes for the stretch and
colour relations (\mbcorr) using a standard empirical approach with the
form:
\begin{equation}
\mbcorr=\mB+\alpha\left(s-1\right)-\beta\col
\label{eq:mbcorr}
\end{equation}
where $\alpha$ parametrizes the stretch--luminosity relation, $\beta$
parametrizes the colour--luminosity relation, and \mB, $s$ and \col\
are the observed peak magnitude, stretch and colour output from the
light curve fitters ($\S$~\ref{sec:sn-light-curve}). \mB\ and \col\
are corrected for Milky Way extinction but not for any host galaxy
extinction. $\alpha$ and $\beta$ are typically determined from
simultaneous fits with the cosmological parameters \citep[e.g.][see
also $\S$~\ref{sec:lum-depend-trends}]{2006A&A...447...31A}, and
reduce the scatter in the peak SN Ia magnitudes to $\sim$0.15\,mag.

\subsection{Light curve shape and colour}
\label{sec:lcs-colour}

The correlation between SN~Ia light curve shape (stretch, \deltam,
etc.) and host galaxy morphology
\citep[e.g.][]{1995AJ....109....1H,1996AJ....112.2398H,1999AJ....117..707R,2000AJ....120.1479H},
SFR \citepalias{2006ApJ...648..868S,2009ApJ...707.1449N}, and
\mstellar\ \citep{2009ApJ...691..661H} is well established: fainter,
lower stretch (high \deltam) SNe~Ia explode in older stellar
populations showing little ongoing star formation.

\begin{figure*}
\centering
\includegraphics[width=0.495\textwidth]{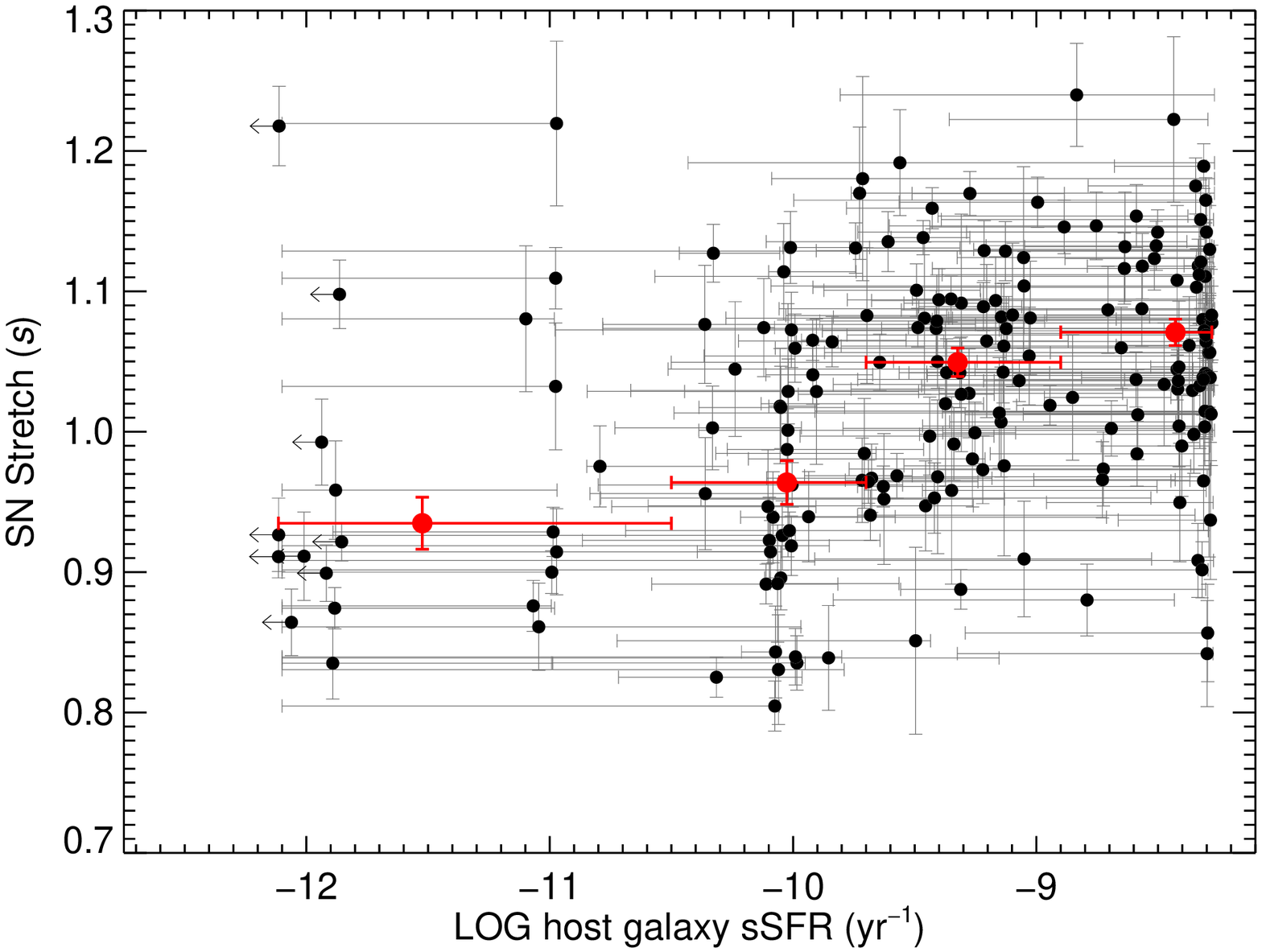}
\includegraphics[width=0.495\textwidth]{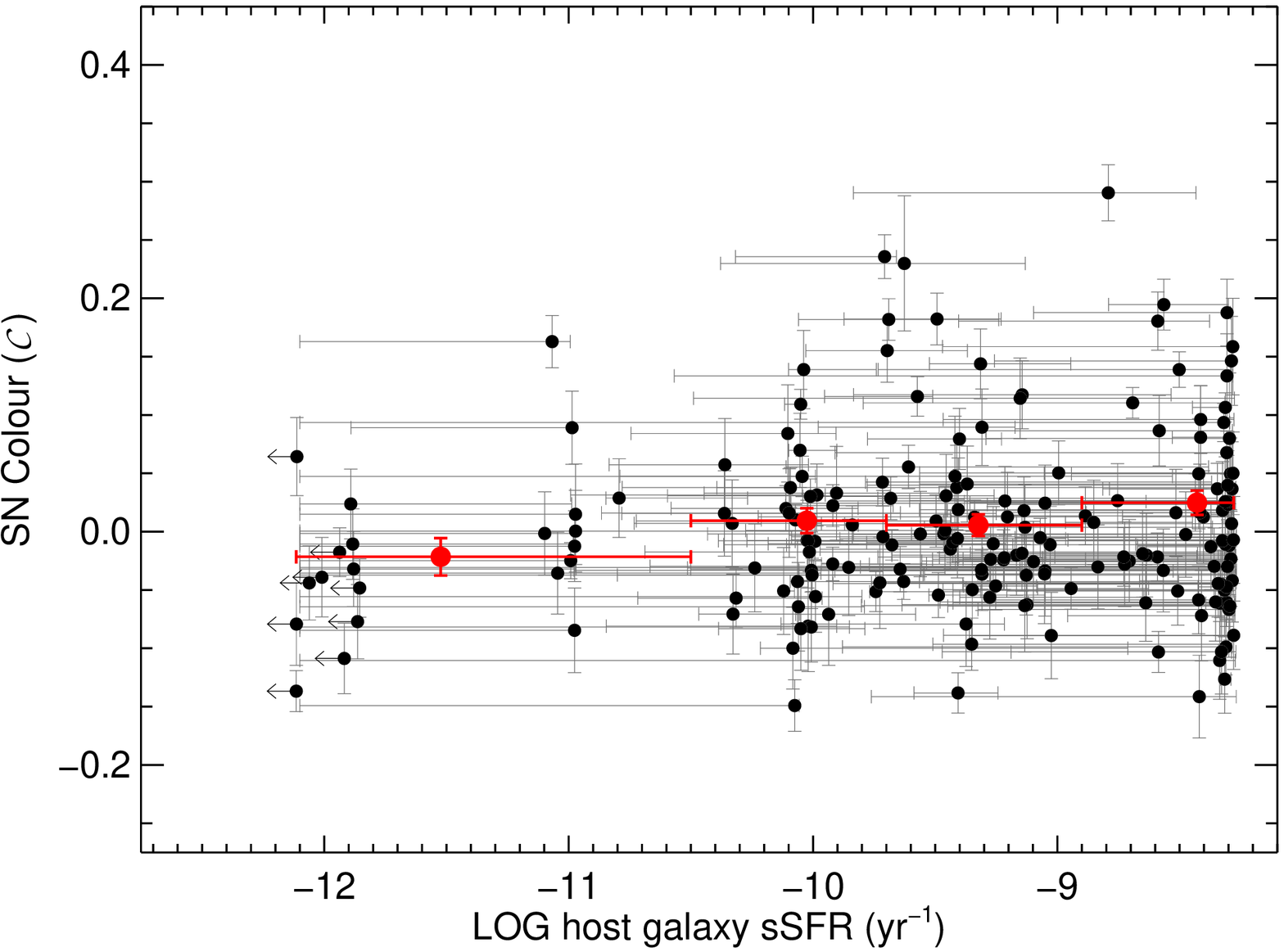}
\includegraphics[width=0.495\textwidth]{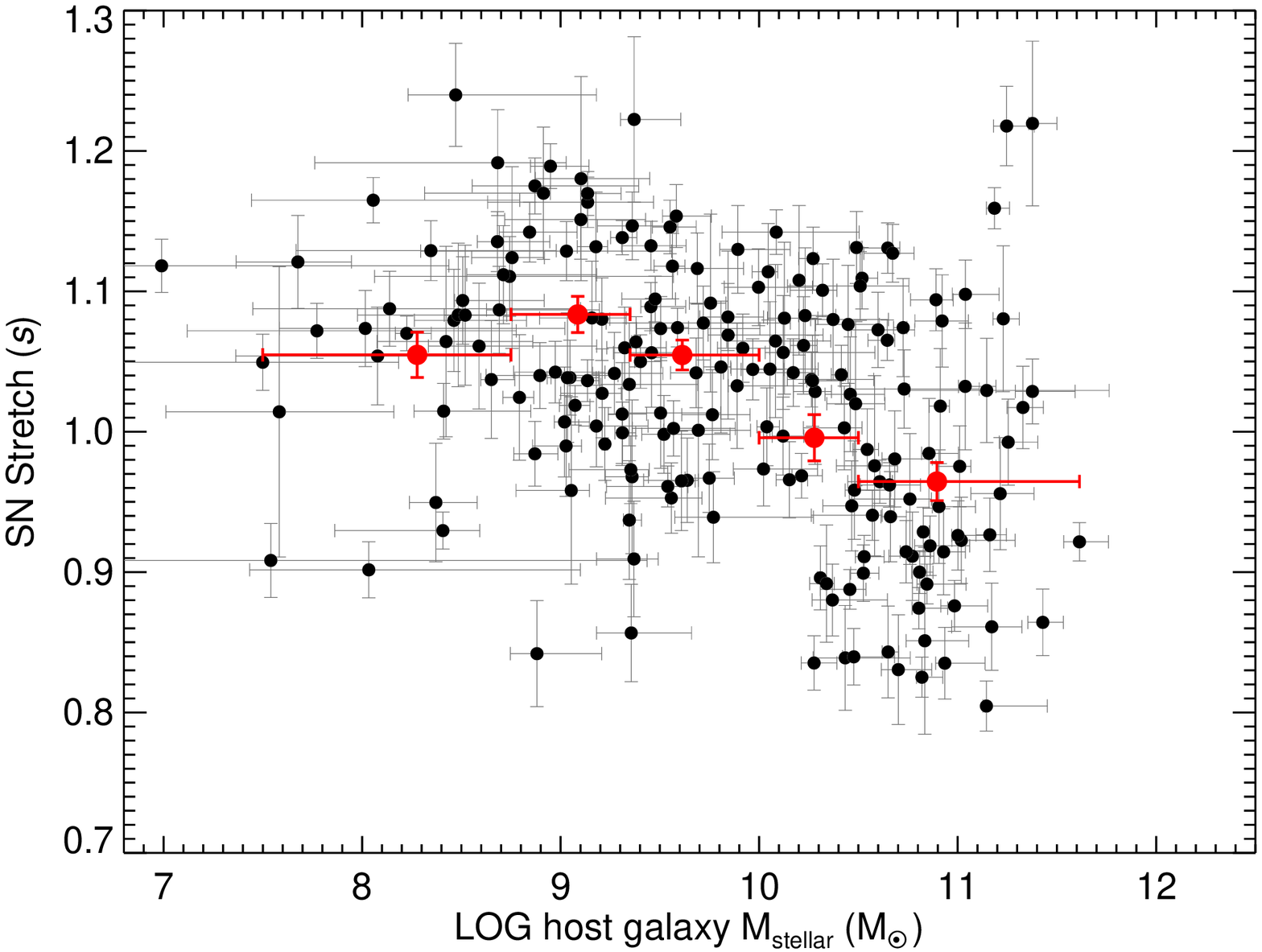}
\includegraphics[width=0.495\textwidth]{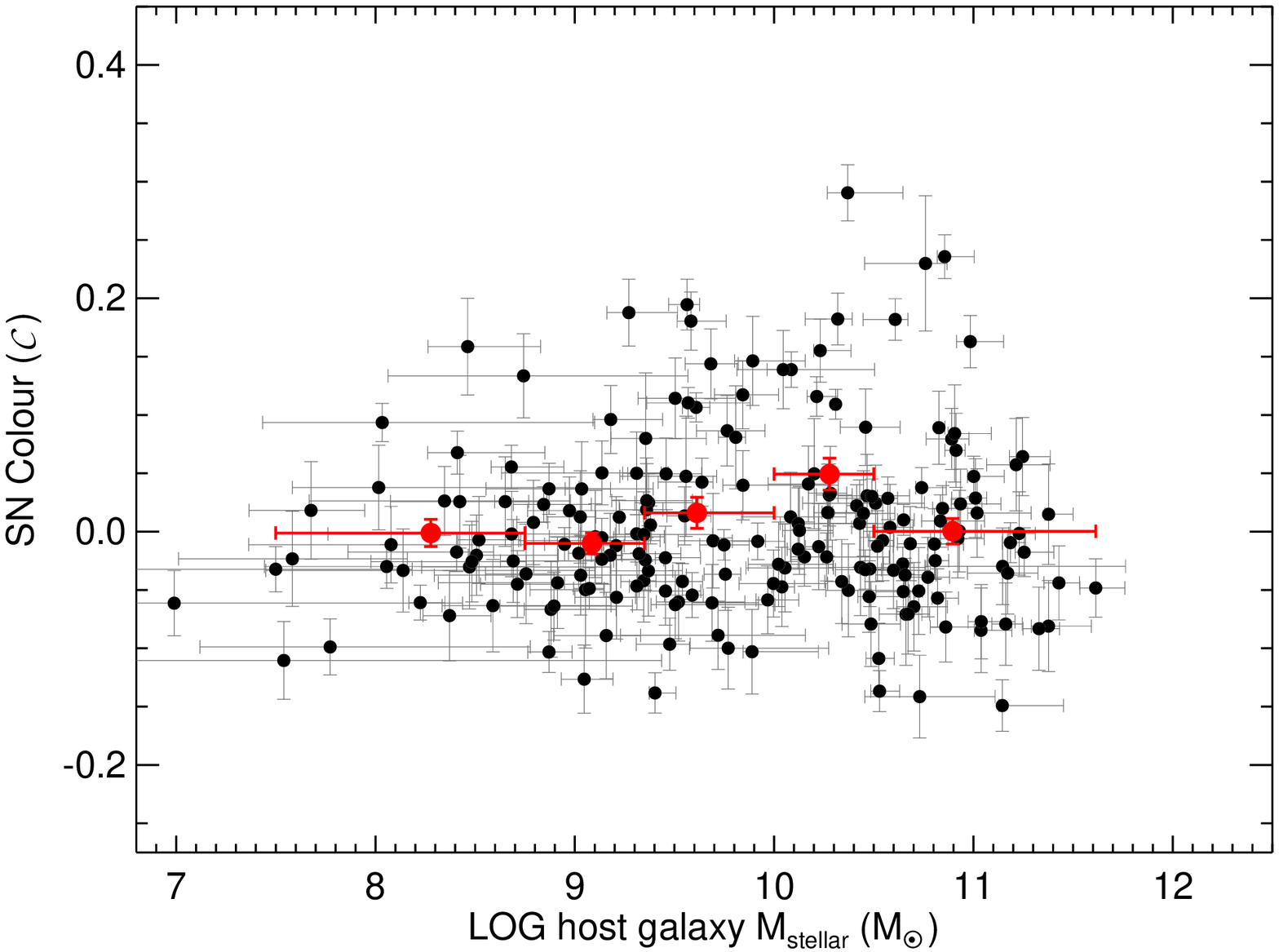}
\caption{The SN~Ia stretch ($s$; left panels) and colour (\col; right
  panels) from SiFTO as a function of the host galaxy sSFR (upper
  panels) and \mstellar\ (lower panels). The red points show the
  weighted mean $s$ or \col, corrected for dispersion, in bins of sSFR
  and \mstellar. Only SNLS SNe are shown (similar plots for the
  low-redshift sample can be found in
  \citetalias{2009ApJ...707.1449N}). Equivalent results are found with
  SALT2.\label{fig:sc_ssfrmass}}
\end{figure*}

We extend this earlier work to examine the SN~Ia properties as a more
continuous function of sSFR and \mstellar\ in
Fig.~\ref{fig:sc_ssfrmass}.  The expected trends with stretch are
recovered. Low stretch SNe Ia preferentially explode in low-sSFR
galaxies with little or no ongoing star formation, and are rarer in
galaxies which have substantial star formation.  By contrast the
higher stretch SNe Ia are found in galaxies with a range of mean sSFRs
but with a deficit in the low sSFR systems. Similar trends are seen as
a function of \mstellar: lower-stretch SNe are almost entirely absent
in low-\mstellar\ galaxies (expected to have the highest sSFRs).
There is is also evidence for stretch being a continuous variable of
sSFR or \mstellar.

No conclusive trends between SN~Ia rest-frame colour and host galaxy
properties have yet been identified. In our data, there are also no
strong trends, though SNLS SNe Ia in low sSFR systems do show slightly
bluer colours in the mean (no differences are seen between the colours
of SNe Ia in low-\mstellar\ hosts and high-\mstellar\ hosts). A
$t$-test shows the mean colours of SNe in low and high sSFR galaxies
have about a 97\% chance of being different; a Mann-Witney $U$ test
gives a similar 97\% chance that the colours arise from different
distributions. This is consistent with the simple viewpoint that more
strongly star-forming galaxies contain more dust, and that dust will
make SN colours redder.  However, the lack of a large difference in
mean colour suggests that the amount of dust along the line of sight
to SNe Ia in star-forming host galaxies is small.

Note that the lack of a trend in the opposite sense may also be
considered surprising at first sight. There is a strong variation in
stretch with star formation activity, and observations show that the
very lowest-$s$ (and typically sub-luminous) SNe Ia are usually redder
in \col\ -- this would imply that the lowest sSFR systems should host
the reddest SNe, which is opposite to the weak trend that we do
observe.  However, the strength of this stretch--colour relationship
for ``normal'' SNe~Ia at maximum light is weak
\citep[e.g.][]{2006AJ....131..527J,2008A&A...487...19N,2008MNRAS.389.1577T}.
Trends do exist between other colours and stretch and at other
light-curve phases
\citep[e.g.][]{1999AJ....118.1766P,2008A&A...487...19N} and for
sub-luminous SN1991bg-like very low-$s$ SNe Ia
\citep{2004ApJ...613.1120G,2006AJ....131..527J}, but these are
excluded from our analysis by the $s>0.8$ requirement.

\section{SN Luminosity dependence on host characteristics}
\label{sec:lum-depend-trends}

We now examine the effect of the environment on the ``corrected''
brightness of SNe Ia, and hence the effect on their use in a
cosmological analysis. We minimize
\begin{equation}
\label{eq:cosmochi2}
\chi^2=\sum_{N}\frac{\left(\mbcorr - \mbmodel\left(z,\scriptm;\omatter\right)\right)^2}{\sigma_{\mathrm{stat}}^2+\sigint^2}
\end{equation}
where \mbcorr\ is given by eqn.~(\ref{eq:mbcorr}),
$\sigma_{\mathrm{stat}}$ is the total identified statistical error and
includes uncertainties in both \mB\ and \mbmodel, \sigint\
parametrizes the intrinsic dispersion of each SN, and the sum is over
the $N$ SNe~Ia entering the fit. \mbmodel\ is the model $B$-band
magnitudes for each SN, given by
\begin{equation}
\label{eq:mtheory}
\mbmodel=5\log_{10}{\mathcal D_L}\left(z;\omatter\right) + \scriptm,
\end{equation}
where ${\mathcal D_L}$ is the $c/H_0$ reduced luminosity distance with
the $c/H_0$ factor (here $c$ is the speed of light) absorbed into
\scriptm, the absolute luminosity of a $s=1$ $\col=0$ SN~Ia
(eqn.~(\ref{eq:mbcorr})). Explicitly,
$\scriptm=\absm+5\log_{10}(c/H_0)+25$, where \absm\ is the absolute
magnitude of a SN~Ia in the $B$-band (for SALT2 fits, \absm\ refers to
an $x_1=0$, $\col=0$ event). For convenience, we present our results
as \absm\ rather than \scriptm, but note that this requires a value of
$H_0$ to be assumed -- a choice that does not impact our results in
any way.

$\alpha$, $\beta$ and \scriptm\ are often referred to as ``nuisance
variables'' in cosmological fits as they are not of immediate interest
when determining cosmological models. Instead they parametrize
luminosity variations within the SN Ia samples and are likely related
to the physics of the SN~Ia explosion and/or the SN~Ia environment;
clearly this makes them of great interest in this paper.

Two different approaches are used. The first approach examines the
residuals of the SNe from global cosmological fits using the SNLS3
$z<0.85$ plus low-redshift sample, fixing $\omatter=0.256$ (the
best-fit value for this sample). We choose this number instead of a
more ``standard'' value like $\omatter=0.3$ to ensure that no redshift
bias in our SN Ia residuals is introduced by adopting a cosmological
model that does not fit the data adequately.  The second examines any
variation of the nuisance variables by fixing the same cosmological
model and performing fits on sub-samples of SNLS SNe with the nuisance
parameters free. The first technique uses global values of the
nuisance variables for the entire sample, whereas the second allows
them to vary by environment. The advantage of the latter technique is
that as the cosmological model is fixed, a large low redshift sample
is not required in order to examine brightness-dependent tests
internally within the well-measured SNLS sample. Throughout, we define
a Hubble residual as $\mbcorr-\mbmodel$, i.e. brighter SNe have
negative Hubble residuals.

\subsection{Residuals from global cosmological fits}
\label{sec:resid-from-glob}

We consider the residuals from the best fitting cosmological model as
a function of three host properties: sSFR, \mstellar, and \mstellar\
converted into a metallicity estimate ($Z$). Residual trends here
indicate luminosity-dependent effects that are not removed by the
standard $s$ (or $x_1$) and \col\ methodology, but that do correlate
with some other physical variable associated with the host galaxy. We
emphasise that \mstellar\ and $Z$ are therefore highly correlated, and
our $Z$ estimates are dependent on the evolution in the \mstellar--$Z$
relation measured by \citet{2009A&A...495...53L}.

\subsubsection{Host specific SFR}
\label{sec:specific-sfr}

\begin{figure*}
\centering
\includegraphics[width=0.75\textwidth]{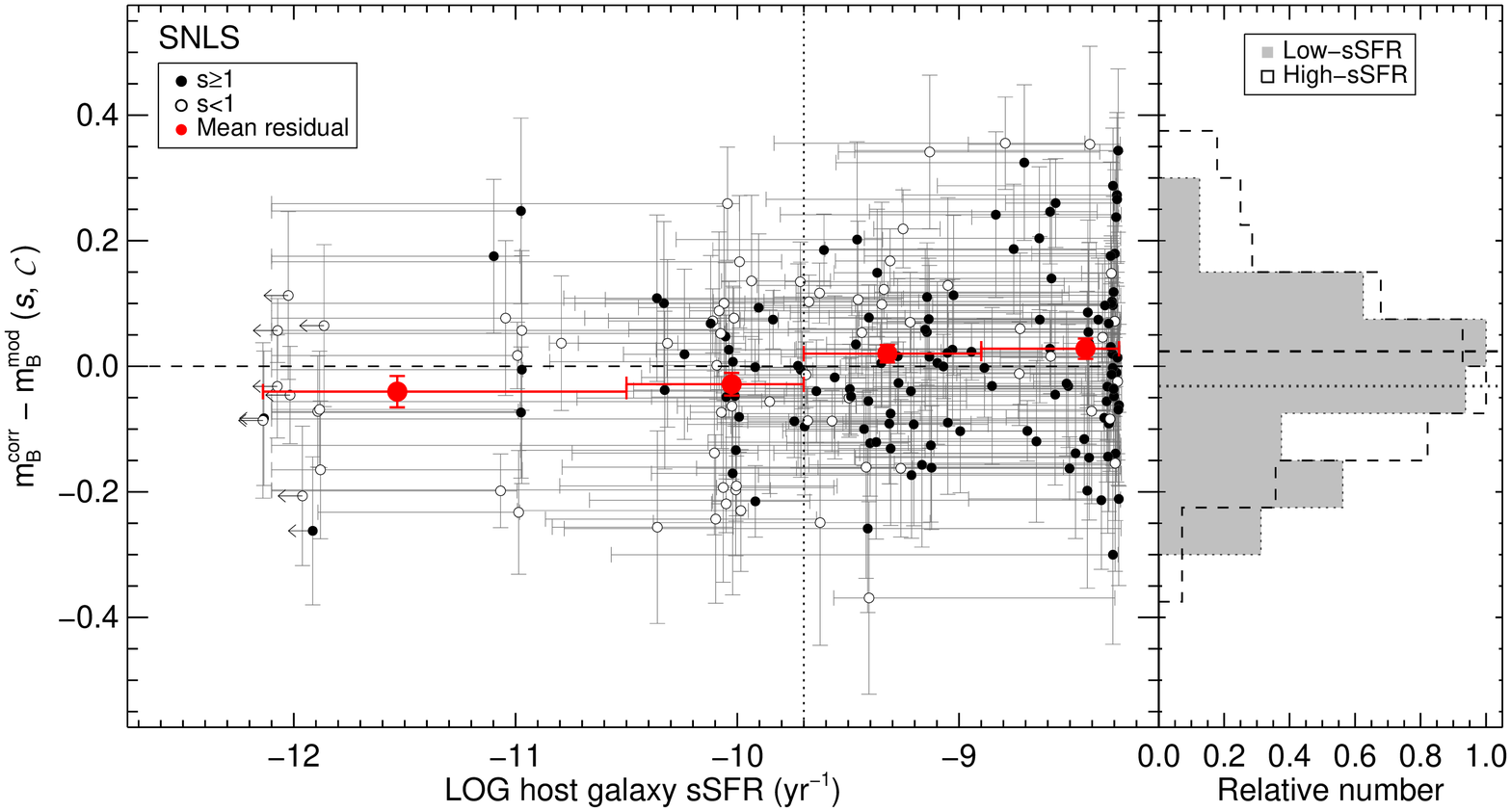}
\includegraphics[width=0.75\textwidth]{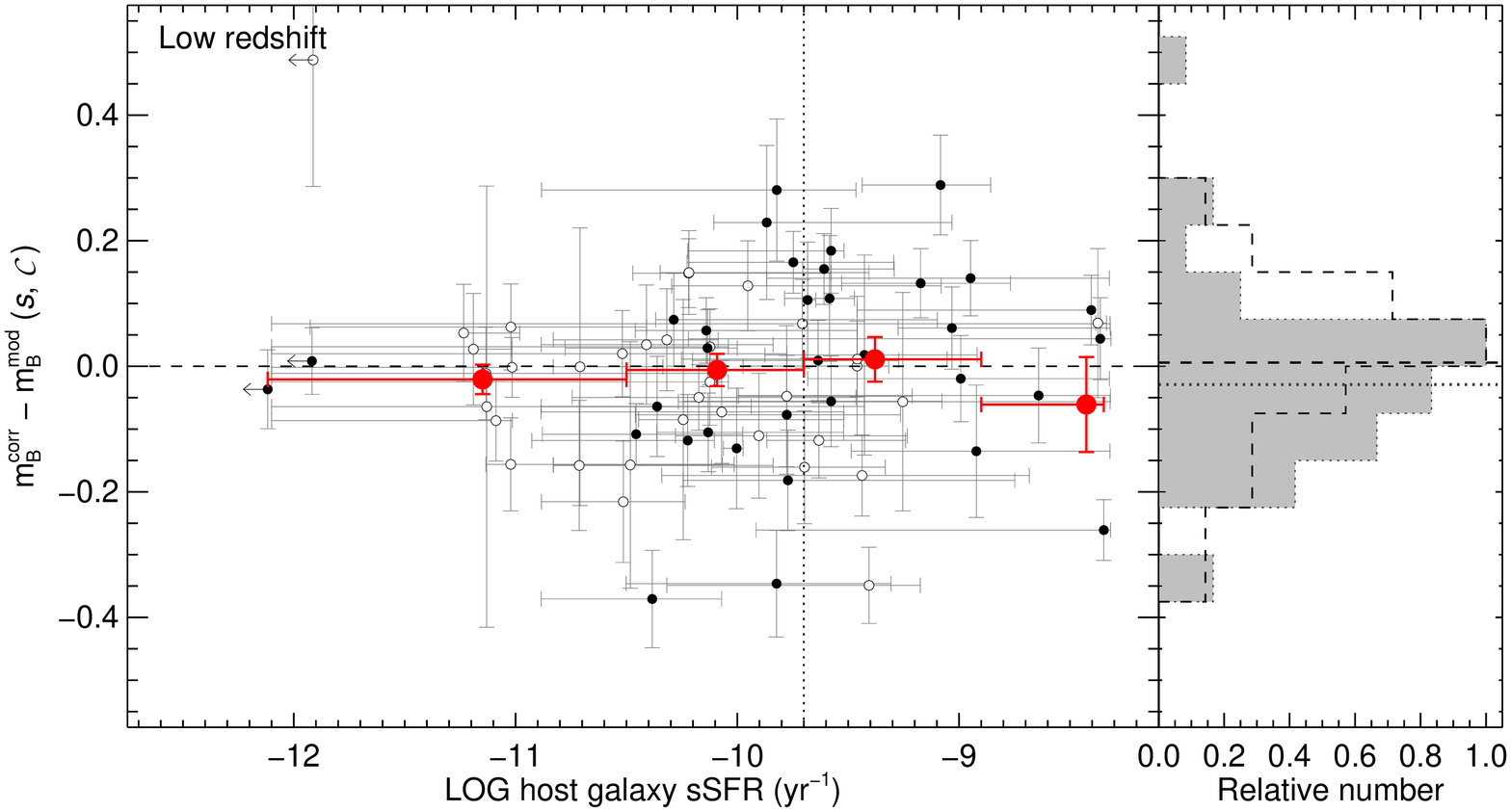}
\caption{Residuals from the best-fitting cosmology for the SNLS (top)
  and low redshift (bottom) samples as a function of host galaxy sSFR.
  Hosts with $s<1$ SNe Ia are shown as open circles, and those with
  $s\geq1$ as filled circles.  Brighter SNe Ia after correction have
  negative residuals.  The error bars on the individual SNe are taken
  from the SED fitting for the sSFR axis, and are the statistical
  errors propagated through the light curve fitting for the residual
  axis (i.e. they do not include the \sigint\ component). The red
  circles are the mean residuals in bins of host sSFR, drawn at the
  mean value of the sSFR in each bin. The error bars on these points
  represent the bin width in sSFR, and the weighted error on the mean
  (corrected for dispersion in each bin) for the residual axis.  The
  right-hand histograms show the total residuals for SNe Ia in low and
  high sSFR hosts.\label{fig:resid_ssfr}}
\end{figure*}

The residuals in the SNLS and low-redshift samples as a function of
host galaxy sSFR are shown Fig.~\ref{fig:resid_ssfr} (the R07 sample
has too few SNe for any meaningful trends to be detected).  In the
SNLS sample, there is evidence that SN Ia events in low sSFR galaxies
appear brighter on average than those in high sSFR galaxies after
$s$/$x_1$ and \col\ corrections.  The numerical values for the mean
residuals in each bin are given in Table~\ref{tab:snlsresids_byssfr}
for SiFTO and Table~\ref{tab:snlsresids_byssfr_salt} for SALT2; the
differences seen are not dependent on the light curve fitter used.
Note that this trend is opposite to that perhaps expected from the
observation that low-stretch (intrinsically fainter) SNe Ia reside in
lower-sSFR hosts ($\S$~\ref{sec:lcs-colour}); however this observation
is pre-stretch correction -- our residuals have been corrected for
both stretch and colour (this is discussed in more detail in
$\S$~\ref{sec:sn-stretch-or}).

\begin{table*}
\centering
\begin{tabular}{ccccccccccc}
  \hline
  Sample & $\log$ sSFR & \multicolumn{3}{c}{All SNe} & \multicolumn{3}{c}{$s<1$} & \multicolumn{3}{c}{$s\geq1$} \\
  & (yr$^{-1}$) & N & Mean residual & Error & N & Mean residual & Error & N & Mean residual & Error\\
  \hline
   SNLS & -11.53 &  22 & -0.040 & 0.025 &  16 & -0.043 & 0.027 &   6 & -0.025 & 0.073\\
        & -10.03 &  43 & -0.029 & 0.018 &  23 & -0.028 & 0.029 &  20 & -0.029 & 0.020\\
        &  -9.32 &  59 &  0.020 & 0.014 &  19 &  0.053 & 0.027 &  40 & -0.005 & 0.014\\
        &  -8.43 &  71 &  0.028 & 0.016 &  12 &  0.084 & 0.042 &  59 &  0.016 & 0.017\\
Low-$z$ & -11.15 &  15 & -0.021 & 0.023\\
        & -10.09 &  29 & -0.006 & 0.026\\
        &  -9.38 &  20 &  0.011 & 0.036\\
        &  -8.43 &   5 & -0.061 & 0.076\\
  \hline
  Sample & $\log$ \mstellar & \multicolumn{3}{c}{All SNe} & \multicolumn{3}{c}{$s<1$} & \multicolumn{3}{c}{$s\geq1$} \\
  & (\msun) & N & Mean residual & Error & N & Mean residual & Error & N & Mean residual & Error\\
  \hline
   SNLS &  8.30 &  27 &  0.041 & 0.025 &   3 &  0.058 & 0.074 &  24 &  0.037 & 0.028\\
        &  9.09 &  35 &  0.021 & 0.015 &   7 &  0.100 & 0.034 &  28 & -0.001 & 0.015\\
        &  9.61 &  40 &  0.048 & 0.018 &  11 &  0.095 & 0.027 &  29 &  0.026 & 0.023\\
        & 10.28 &  36 & -0.023 & 0.025 &  13 & -0.007 & 0.050 &  23 & -0.038 & 0.025\\
        & 10.90 &  56 & -0.037 & 0.014 &  36 & -0.032 & 0.019 &  20 & -0.049 & 0.021\\
Low-$z$ &  7.88 &   2 &    ... &    ...\\
        &  9.01 &   3 &  0.115 & 0.031\\
        &  9.79 &   6 &  0.090 & 0.052\\
        & 10.32 &  20 &  0.018 & 0.030\\
        & 10.94 &  38 & -0.028 & 0.021\\
  \hline
  Sample & $Z$ & \multicolumn{3}{c}{All SNe} & \multicolumn{3}{c}{$s<1$} & \multicolumn{3}{c}{$s\geq1$} \\
  & (12+$\log(\mathrm{O}/\mathrm{H})$) & N & Mean residual & Error & N & Mean residual & Error & N & Mean residual & Error\\
  \hline
   SNLS &  8.24 &  41 &  0.023 & 0.018 &   7 &  0.042 & 0.035 &  34 &  0.018 & 0.021\\
        &  8.56 &  35 &  0.020 & 0.018 &   8 &  0.070 & 0.060 &  27 &  0.011 & 0.018\\
        &  8.72 &  35 &  0.059 & 0.019 &   8 &  0.106 & 0.024 &  27 &  0.035 & 0.024\\
        &  8.88 &  66 & -0.034 & 0.017 &  35 & -0.020 & 0.026 &  31 & -0.056 & 0.020\\
        &  9.00 &  15 & -0.022 & 0.022 &  12 & -0.012 & 0.027 &   3 & -0.051 & 0.033\\
Low-$z$ &  8.07 &   2 &    ... &    ...\\
        &  8.55 &   2 &    ... &    ...\\
        &  8.73 &   2 &    ... &    ...\\
        &  8.91 &   5 &  0.049 & 0.039\\
        &  9.06 &  58 & -0.015 & 0.017\\
  \hline
\end{tabular}
\caption{Binned SN Ia residuals for the SNLS and low-redshift samples as a function of host galaxy sSFR, \mstellar, and $Z$, calculated with the SiFTO light curve fitter. The host parameter bin centres are given as the mean of all the hosts in that bin. The mean residuals are the weighted average of SNe in each bin, and the errors are the errors on that weighted mean.\label{tab:snlsresids_byssfr}}
\end{table*}

\begin{table*}
\centering
\begin{tabular}{ccccccccccc}
  \hline
  Sample & $\log$ sSFR & \multicolumn{3}{c}{All SNe} & \multicolumn{3}{c}{$x_1<0$} & \multicolumn{3}{c}{$x_1\geq0$} \\
  & (yr$^{-1}$) & N & Mean residual & Error & N & Mean residual & Error & N & Mean residual & Error\\
  \hline
   SNLS & -11.51 &  21 & -0.036 & 0.029 &  15 & -0.045 & 0.029 &   6 &  0.009 & 0.093\\
        & -10.03 &  37 & -0.021 & 0.022 &  20 & -0.023 & 0.037 &  17 & -0.020 & 0.021\\
        &  -9.31 &  61 &  0.027 & 0.014 &  21 &  0.042 & 0.023 &  40 &  0.016 & 0.018\\
        &  -8.39 &  70 &  0.027 & 0.017 &  14 & -0.002 & 0.040 &  56 &  0.034 & 0.019\\
Low-$z$ & -11.36 &  21 &  0.017 & 0.031\\
        & -10.06 &  31 & -0.007 & 0.026\\
        &  -9.38 &  20 &  0.026 & 0.035\\
        &  -8.42 &   6 & -0.092 & 0.076\\
  \hline
  Sample & $\log$ \mstellar & \multicolumn{3}{c}{All SNe} & \multicolumn{3}{c}{$x_1<0$} & \multicolumn{3}{c}{$x_1\geq0$} \\
  & (\msun) & N & Mean residual & Error & N & Mean residual & Error & N & Mean residual & Error\\
  \hline
   SNLS &  8.31 &  25 &  0.042 & 0.030 &   4 &  0.041 & 0.069 &  21 &  0.042 & 0.035\\
        &  9.08 &  32 &  0.027 & 0.019 &   8 &  0.068 & 0.035 &  24 &  0.012 & 0.022\\
        &  9.61 &  39 &  0.063 & 0.016 &  11 &  0.042 & 0.029 &  28 &  0.073 & 0.020\\
        & 10.27 &  32 & -0.017 & 0.024 &  12 & -0.006 & 0.045 &  20 & -0.026 & 0.026\\
        & 10.89 &  60 & -0.036 & 0.018 &  35 & -0.031 & 0.024 &  25 & -0.045 & 0.027\\
Low-$z$ &  7.88 &   2 &    ... &    ...\\
        &  9.01 &   3 &  0.074 & 0.042\\
        &  9.81 &   7 &  0.038 & 0.063\\
        & 10.30 &  21 &  0.021 & 0.029\\
        & 10.95 &  45 & -0.009 & 0.023\\
  \hline
  Sample & $Z$ & \multicolumn{3}{c}{All SNe} & \multicolumn{3}{c}{$x_1<0$} & \multicolumn{3}{c}{$x_1\geq0$} \\
  & (12+$\log(\mathrm{O}/\mathrm{H})$) & N & Mean residual & Error & N & Mean residual & Error & N & Mean residual & Error\\
  \hline
   SNLS &  8.25 &  38 &  0.023 & 0.024 &   7 &  0.043 & 0.042 &  31 &  0.016 & 0.029\\
        &  8.55 &  30 &  0.023 & 0.019 &   8 & -0.047 & 0.032 &  22 &  0.037 & 0.022\\
        &  8.72 &  37 &  0.069 & 0.017 &  10 &  0.064 & 0.034 &  27 &  0.071 & 0.019\\
        &  8.88 &  65 & -0.023 & 0.018 &  32 & -0.014 & 0.028 &  33 & -0.036 & 0.023\\
        &  9.00 &  16 & -0.037 & 0.028 &  12 & -0.020 & 0.035 &   4 & -0.084 & 0.040\\
Low-$z$ &  8.07 &   2 &    ... &    ...\\
        &  8.55 &   2 &    ... &    ...\\
        &  8.73 &   2 &    ... &    ...\\
        &  8.92 &   6 & -0.010 & 0.062\\
        &  9.07 &  66 & -0.000 & 0.018\\
  \hline
\end{tabular}
\caption{As Table~\ref{tab:snlsresids_byssfr}, but for the SALT2 light curve fitter. Note that the number of SNe Ia in each bin can vary from that in Table~\ref{tab:snlsresids_byssfr} due to slightly different SN parameter culls for SiFTO and SALT2.\label{tab:snlsresids_byssfr_salt}}
\end{table*}

The differences in the SNLS samples between the two lowest and two
highest sSFR bins are $\simeq$2.6$\sigma$ significance.  In the
low-redshift sample analysed here, the trends are consistent with the
SNLS sample (discarding the most strongly star-forming bin with only a
small number of events), but low significance. For the SNLS data,
fitting a straight line to the binned points and accounting for errors
in both axes detects a non-zero gradient at $\simeq$2.5$\sigma$.

\subsubsection{Host stellar mass}
\label{sec:stellar-mass}

\begin{figure*}
\centering
\includegraphics[width=0.75\textwidth]{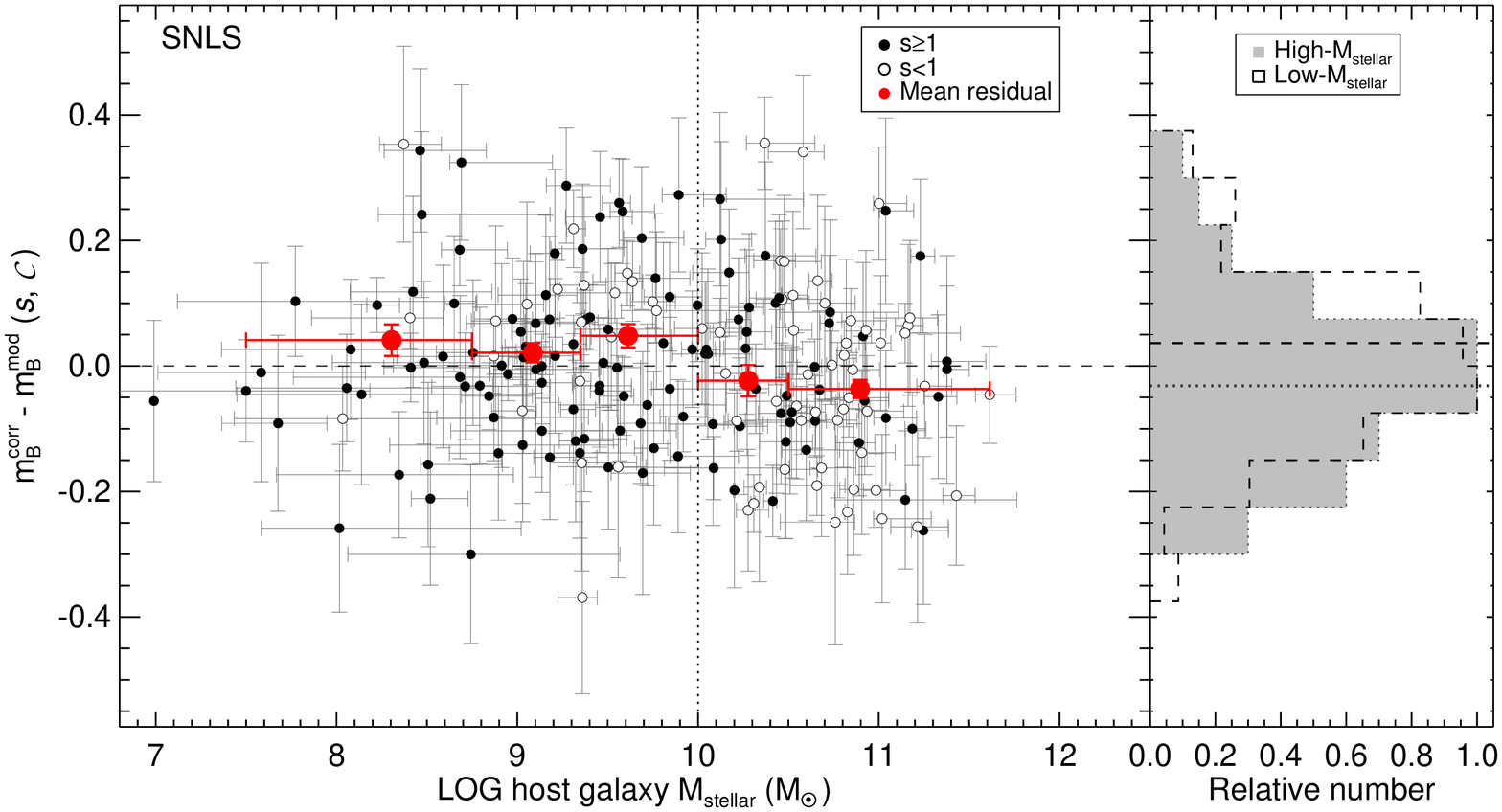}
\includegraphics[width=0.75\textwidth]{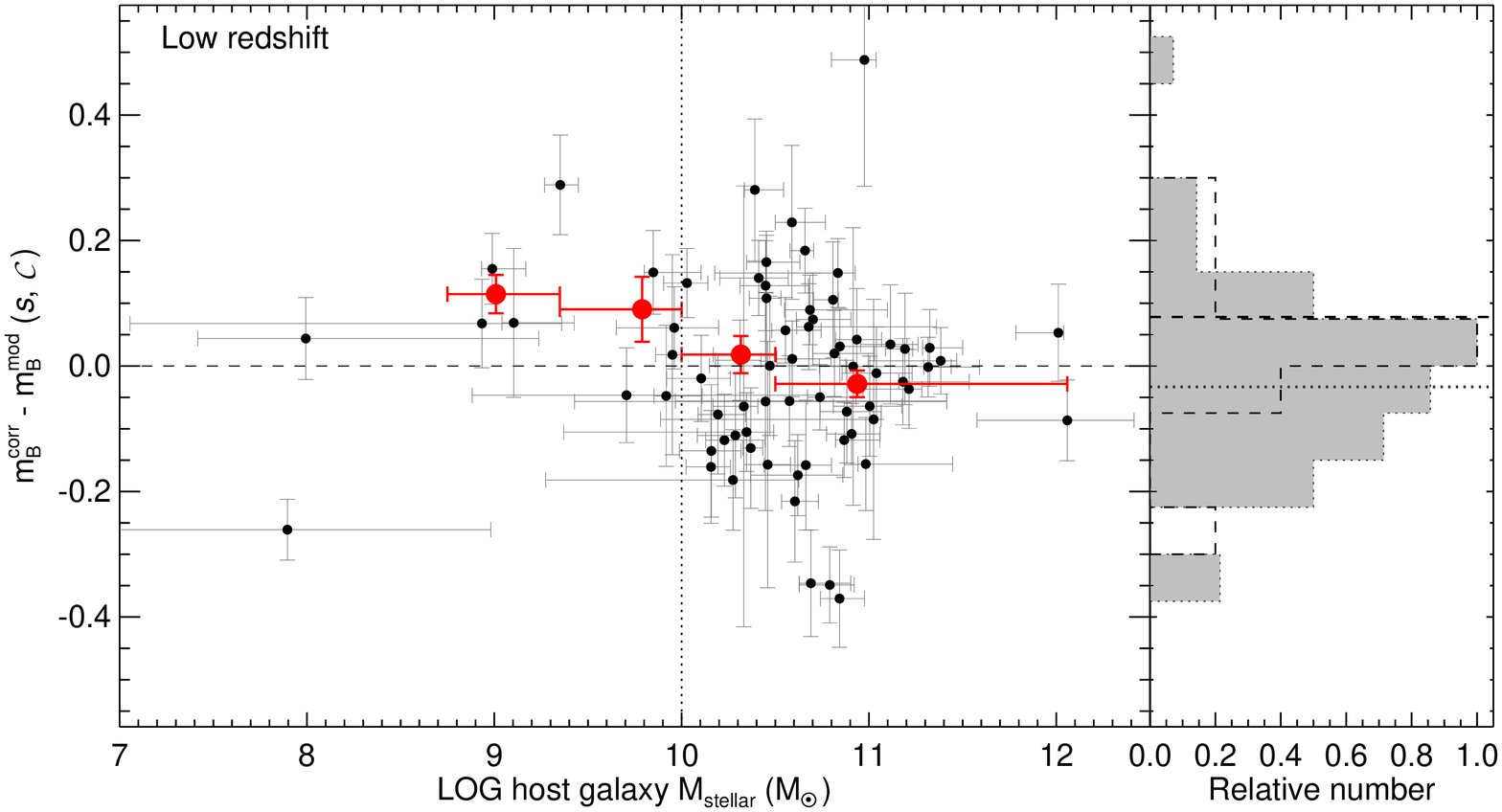}
\caption{As Fig.\ref{fig:resid_ssfr}, but for \mstellar\ instead of sSFR.\label{fig:resid_mass}}
\end{figure*}

The residuals as a function of host galaxy \mstellar\ are shown in
Fig.~\ref{fig:resid_mass}.  In this case, SNe in more massive galaxies
appear brighter on average than those in lower-mass galaxies, after
$s$/$x_1$ and \col\ corrections. The mean residuals are given in
Tables~\ref{tab:snlsresids_byssfr}
and~\ref{tab:snlsresids_byssfr_salt}, middle panel. In the SNLS
sample, SNe in the three lowest \mstellar\ bins are fainter than those
in the two highest \mstellar\ bins at $\simeq$3.9$\sigma$
significance.  The low-redshift sample is consistent with this, though
again the significance is smaller and the range in \mstellar\ is much
more restricted.  For the SNLS data, fitting a straight line to the
binned points detects a non-zero gradient at $\simeq$3.1$\sigma$.

\subsubsection{Host inferred metallicity}
\label{sec:host-inferrr-metall}

\begin{figure*}
\centering
\includegraphics[width=0.75\textwidth]{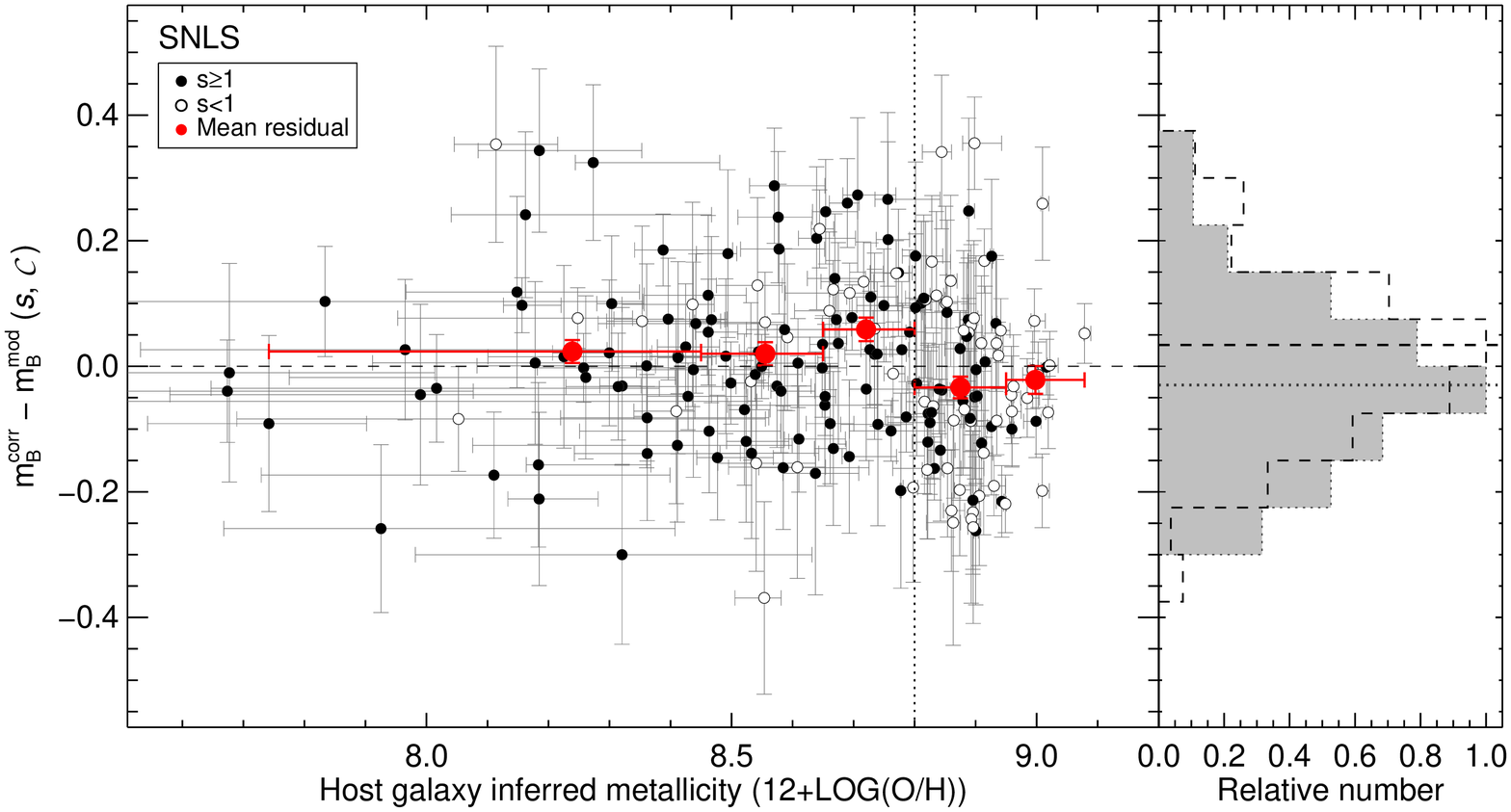}
\includegraphics[width=0.75\textwidth]{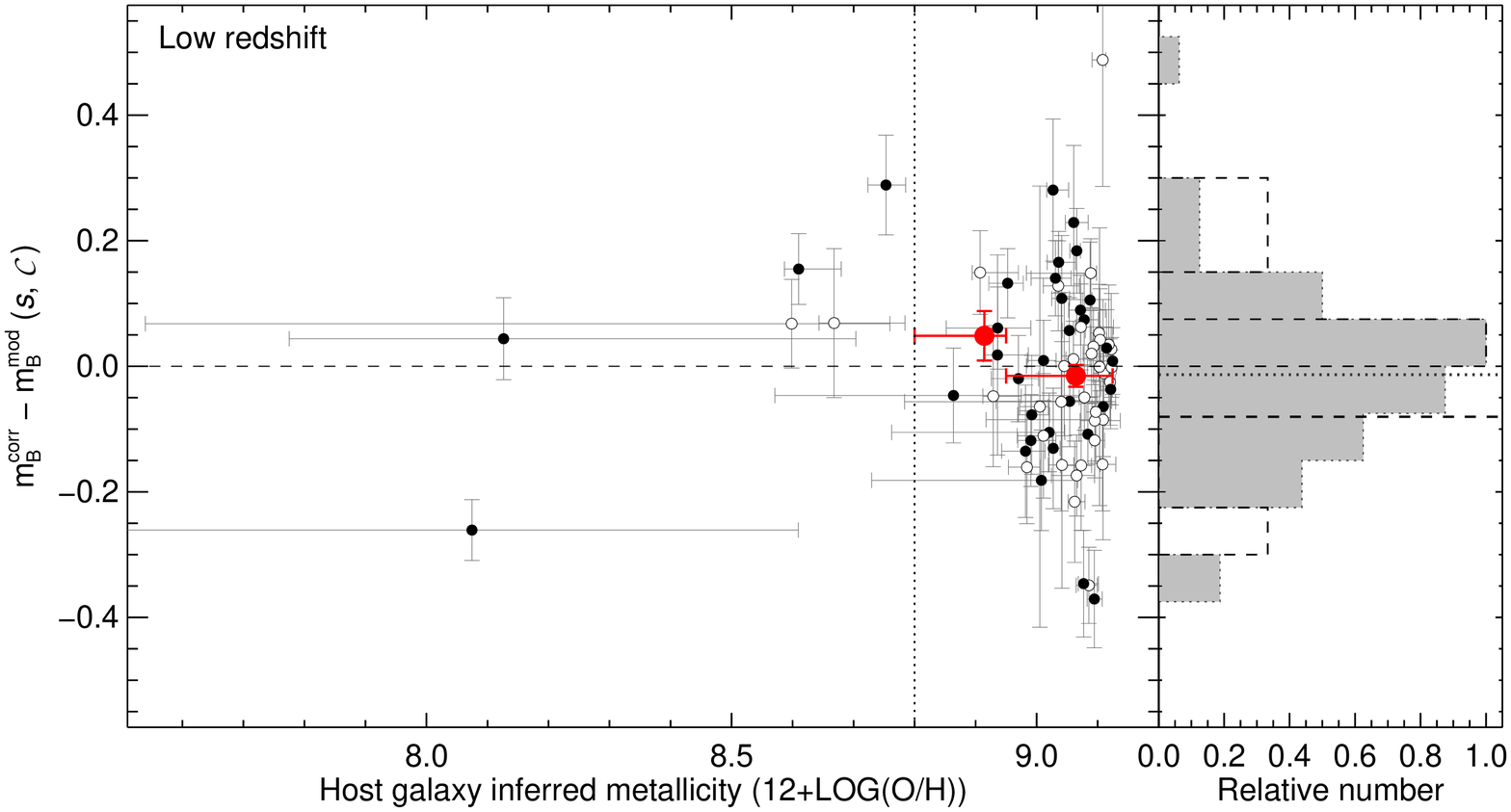}
\caption{As Fig.\ref{fig:resid_ssfr}, but for metallicity instead of
  sSFR.\label{fig:resid_metal}}
\end{figure*}

We convert our \mstellar\ into gas-phase metallicities ($Z$) as
described in $\S$~\ref{sec:galaxy-param-estim}. The residuals as a
function of host galaxy $Z$ are shown in Fig.~\ref{fig:resid_metal}.
As the $Z$ estimates are directly related to the \mstellar, the same
trends are apparent -- indeed, had we used an \mstellar--$Z$ relation
that did not evolve with redshift
\citep[e.g.][]{2004ApJ...613..898T,2005MNRAS.362...41G}, the $Z$ and
\mstellar\ results would be identical. The mean residuals are given in
Table~\ref{tab:snlsresids_byssfr}
and~\ref{tab:snlsresids_byssfr_salt}, lower panel. In the SNLS sample,
SNe in low-$Z$ galaxies are fainter than those in high-$Z$ galaxies at
$\simeq$3.7$\sigma$ significance.  The low-redshift sample is
consistent with this, though the range in $Z$ is more restricted.
Fitting a straight line to the binned SNLS points detects a non-zero
gradient at $\simeq$3.0$\sigma$.

\subsection{Universality of the nuisance variables}
\label{sec:test-univ-nuis}

We now consider the $s$/$x_1$ and \col\ corrections that are applied
to the SN magnitudes, and test whether these two corrections are
consistent for sub-samples of SNe divided according to the
characteristics of their host galaxies. We perform these tests using
the larger and more complete SNLS SN Ia sample, and again hold the
cosmological model fixed.

As discussed in Section~\ref{sec:galaxy-param-estim}, we separate the
hosts according to their sSFR, \mstellar, and $Z$. The exact values
chosen as the split points are a somewhat subjective choice (e.g.
Fig.~\ref{fig:sfrmass}). The \mstellar\ split point was chosen to
separate the hosts into bins of approximately equal sizes, and the $Z$
split point is this \mstellar\ value converted into a $Z$ at $z=0.5$;
no fine-tuning was attempted.  We explore how our results change as we
modify the default splits of $\S$~\ref{sec:galaxy-param-estim} over a
range of 0.8 in $\log(\mathrm{sSFR})$, 1.0 dex in $\log(\mstellar)$,
and 0.2 dex in $Z$. We perform fits for $\alpha$, $\beta$ and
\scriptm, in each case iteratively adjusting \sigint\ until a $\chi^2$
per degree of freedom of 1.0 is obtained.  The SiFTO results are given
in Tables~\ref{tab:alphabeta_bysfr_sifto},
\ref{tab:alphabeta_bymass_sifto}, and
\ref{tab:alphabeta_bymetal_sifto}.

\begin{table*}
\centering
\begin{tabular}{ccccccccccc}
\hline
LOG sSFR  & \multicolumn{5}{c}{Low sSFR hosts} & \multicolumn{5}{c}{High sSFR hosts}\\
split (yr$^{-1}$) & $N_{\mathrm{SN}}$ & $\alpha$ & $\beta$ & \absm & r.m.s. & $N_{\mathrm{SN}}$ & $\alpha$ & $\beta$ & \absm & r.m.s. \\
\hline
 -9.30 &  96 & 1.376$\pm$0.128 & 2.876$\pm$0.159 & -19.193$\pm$0.012 & 0.127 & 100 & 1.757$\pm$0.181 & 3.729$\pm$0.162 & -19.126$\pm$0.017 & 0.143\\
 -9.50 &  74 & 1.305$\pm$0.148 & 2.928$\pm$0.184 & -19.196$\pm$0.015 & 0.123 & 122 & 1.771$\pm$0.161 & 3.567$\pm$0.148 & -19.130$\pm$0.015 & 0.151\\
 -9.70 &  61 & 1.379$\pm$0.148 & 2.728$\pm$0.231 & -19.208$\pm$0.016 & 0.120 & 135 & 1.576$\pm$0.149 & 3.445$\pm$0.136 & -19.149$\pm$0.014 & 0.147\\
 -9.90 &  53 & 1.338$\pm$0.170 & 2.719$\pm$0.232 & -19.204$\pm$0.018 & 0.123 & 143 & 1.576$\pm$0.145 & 3.455$\pm$0.136 & -19.152$\pm$0.013 & 0.145\\
-10.10 &  27 & 1.582$\pm$0.255 & 2.976$\pm$0.451 & -19.205$\pm$0.026 & 0.128 & 169 & 1.394$\pm$0.114 & 3.372$\pm$0.125 & -19.167$\pm$0.011 & 0.142\\
\hline
\end{tabular}
\caption{SiFTO fits with a fixed cosmological model for low and high sSFR  galaxies with different values of sSFR (given in column one) used to split the sample. The nuisance parameters are free in the fit.\label{tab:alphabeta_bysfr_sifto}}
\end{table*}

\begin{table*}
\centering
\begin{tabular}{ccccccccccc}
\hline
LOG \mstellar & \multicolumn{5}{c}{High-\mstellar\ hosts} & \multicolumn{5}{c}{Low-\mstellar\ hosts}\\
split (\msun)& $N_{\mathrm{SN}}$ & $\alpha$ & $\beta$ & \absm & r.m.s. & $N_{\mathrm{SN}}$ & $\alpha$ & $\beta$ & \absm & r.m.s. \\
\hline
 9.50 & 120 & 1.443$\pm$0.138 & 3.380$\pm$0.143 & -19.189$\pm$0.013 & 0.142 &  75 & 1.660$\pm$0.189 & 3.601$\pm$0.250 & -19.117$\pm$0.020 & 0.147\\
 9.75 & 103 & 1.501$\pm$0.143 & 3.258$\pm$0.154 & -19.199$\pm$0.014 & 0.141 &  92 & 1.612$\pm$0.173 & 3.644$\pm$0.188 & -19.122$\pm$0.017 & 0.146\\
10.00 &  92 & 1.554$\pm$0.148 & 3.159$\pm$0.163 & -19.206$\pm$0.014 & 0.141 & 103 & 1.638$\pm$0.164 & 3.714$\pm$0.170 & -19.121$\pm$0.016 & 0.143\\
10.25 &  77 & 1.555$\pm$0.162 & 3.203$\pm$0.180 & -19.213$\pm$0.016 & 0.143 & 118 & 1.673$\pm$0.159 & 3.499$\pm$0.156 & -19.127$\pm$0.015 & 0.141\\
10.50 &  56 & 1.577$\pm$0.165 & 2.937$\pm$0.195 & -19.222$\pm$0.017 & 0.131 & 139 & 1.515$\pm$0.141 & 3.480$\pm$0.145 & -19.148$\pm$0.013 & 0.142\\
\hline
\end{tabular}
\caption{As Table~\ref{tab:alphabeta_bysfr_sifto}, but for \mstellar\ instead of sSFR.\label{tab:alphabeta_bymass_sifto}}
\end{table*}

\begin{table*}
\centering
\begin{tabular}{ccccccccccc}
\hline
$Z$ split & \multicolumn{5}{c}{High-$Z$ hosts} & \multicolumn{5}{c}{Low-$Z$ hosts}\\
(12+$\log(\mathrm{O}/\mathrm{H})$)& $N_{\mathrm{SN}}$ & $\alpha$ & $\beta$ & \absm & r.m.s. & $N_{\mathrm{SN}}$ & $\alpha$ & $\beta$ & \absm & r.m.s. \\
\hline
 8.70 & 102 & 1.512$\pm$0.147 & 3.238$\pm$0.158 & -19.195$\pm$0.014 & 0.142 &  93 & 1.661$\pm$0.175 & 3.800$\pm$0.192 & -19.115$\pm$0.018 & 0.149\\
 8.75 &  92 & 1.577$\pm$0.152 & 3.172$\pm$0.165 & -19.204$\pm$0.015 & 0.143 & 103 & 1.688$\pm$0.161 & 3.848$\pm$0.176 & -19.110$\pm$0.016 & 0.146\\
 8.80 &  81 & 1.601$\pm$0.155 & 3.102$\pm$0.170 & -19.210$\pm$0.015 & 0.139 & 114 & 1.612$\pm$0.160 & 3.747$\pm$0.165 & -19.124$\pm$0.015 & 0.144\\
 8.85 &  61 & 1.534$\pm$0.185 & 3.095$\pm$0.194 & -19.217$\pm$0.018 & 0.139 & 134 & 1.739$\pm$0.148 & 3.614$\pm$0.151 & -19.125$\pm$0.014 & 0.143\\
 8.90 &  34 & 1.526$\pm$0.187 & 2.795$\pm$0.225 & -19.218$\pm$0.021 & 0.113 & 161 & 1.491$\pm$0.130 & 3.619$\pm$0.139 & -19.151$\pm$0.012 & 0.148\\
\hline
\end{tabular}
\caption{As Table~\ref{tab:alphabeta_bysfr_sifto}, but for $Z$ instead of sSFR.\label{tab:alphabeta_bymetal_sifto}}
\end{table*}

\begin{figure*}
\centering
\includegraphics[width=0.495\textwidth]{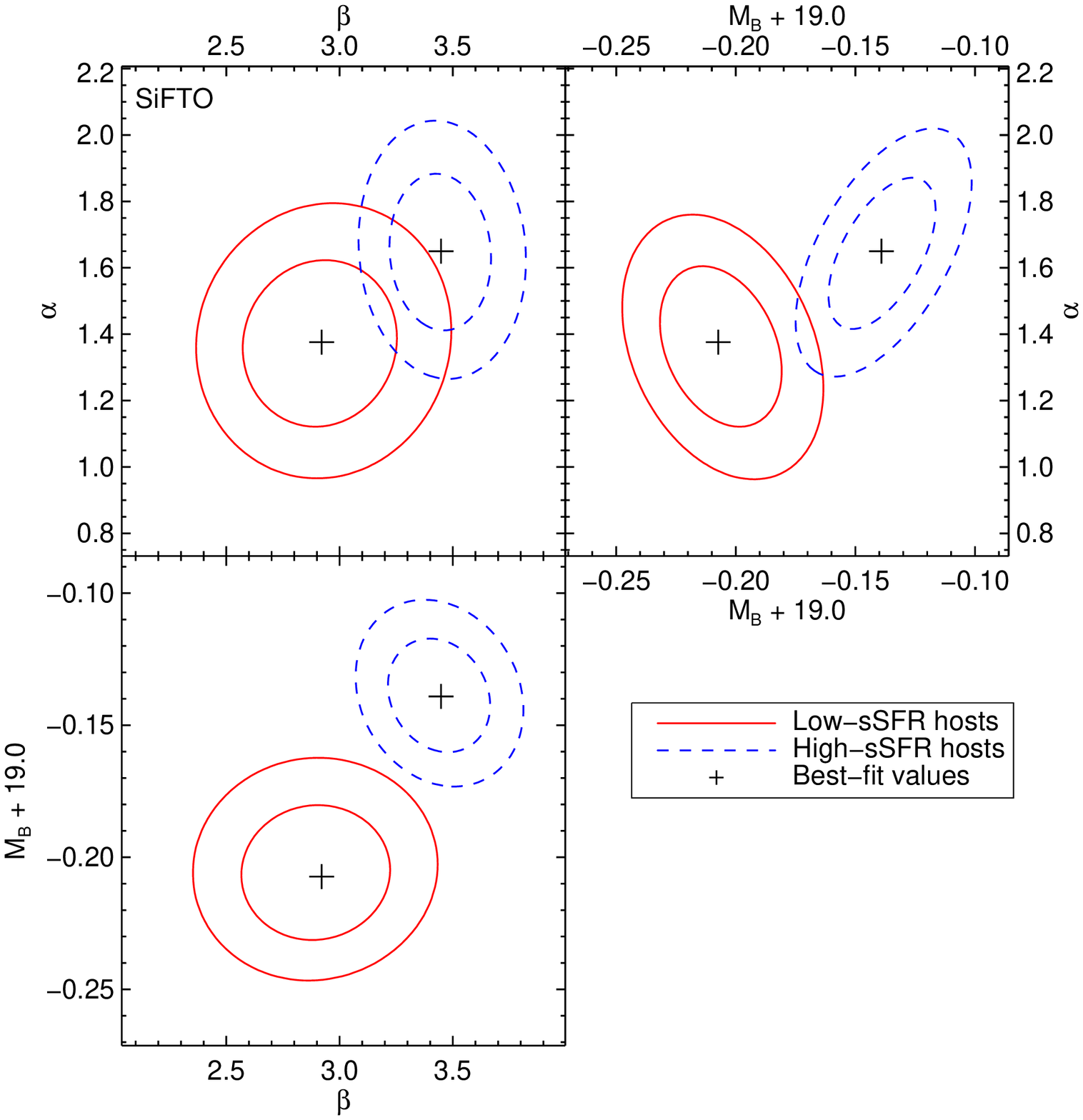}
\includegraphics[width=0.495\textwidth]{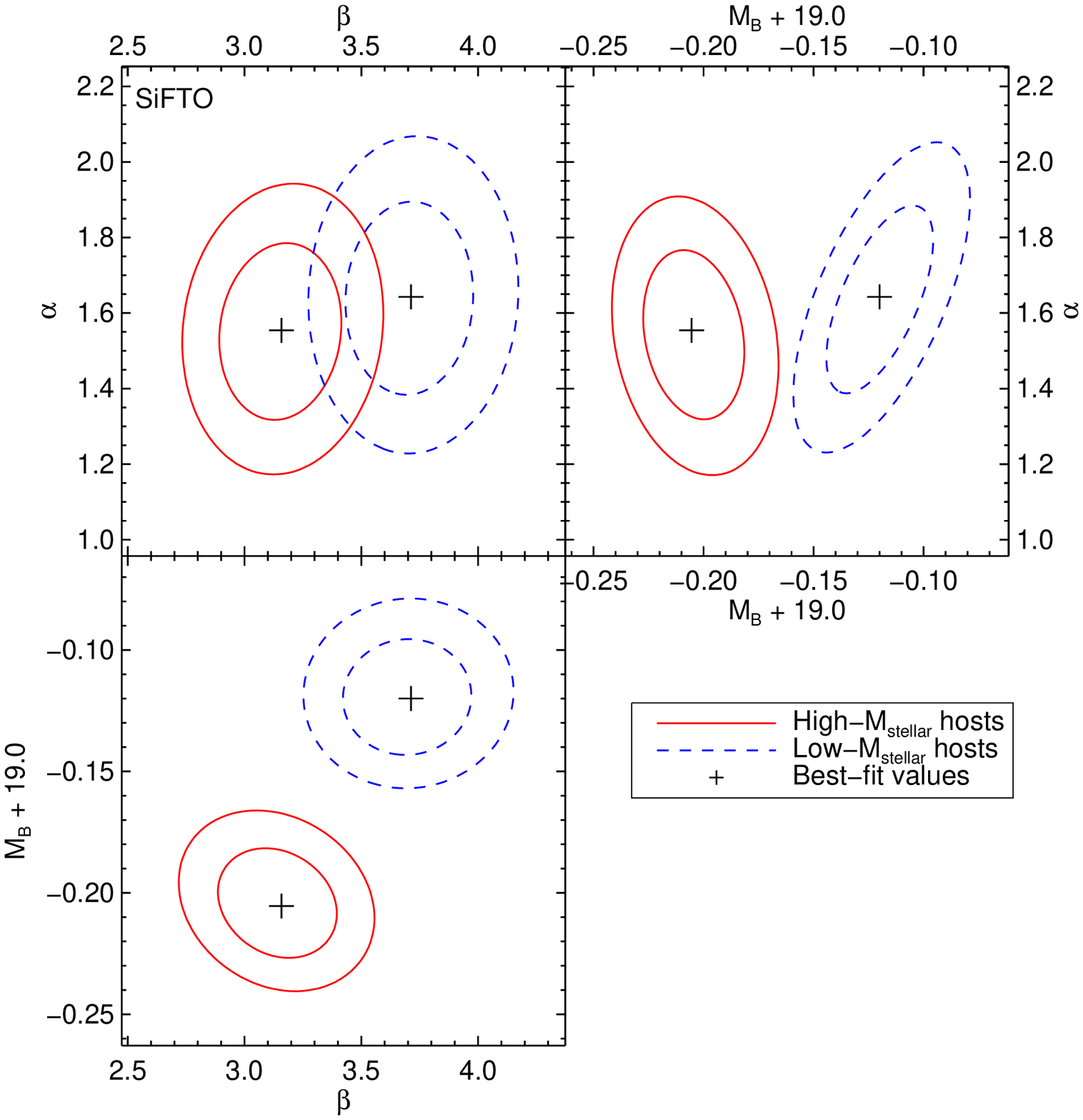}
\caption{Joint confidence contours (one and two $\sigma$) in $\alpha$
  and $\beta$ (top left), $\alpha$ and \absm\ (top right), and \absm\
  and $\beta$ (lower left) for fits where the SNe are split according
  to host galaxy sSFR (left panels) and \mstellar\ (right panels).
  Plots with SNe split by $Z$ are similar to those for \mstellar\ and
  are not shown. The crosses show the best fit
  values.\label{fig:abbyssfrmass}}
\end{figure*}

As expected from $\S$~\ref{sec:resid-from-glob}, there are several
host dependent trends.  In particular, we find that
\begin{itemize}
\item SNe Ia in high-\mstellar\ or high-$Z$ hosts prefer brighter \absm\
  than those in low-\mstellar\ (low-$Z$) hosts ($\simeq4.0\sigma$), as
  do SNe in low-sSFR versus high-sSFR hosts ($\simeq2.5\sigma$), both
  in the same sense as the trends in Fig.~\ref{fig:resid_ssfr},
\item SNe Ia in low-sSFR hosts show smaller values of $\beta$ than
  those in high-sSFR hosts ($\simeq2.7\sigma$). Smaller ($<2.5\sigma$)
  differences are also seen between $\beta$ measured in high and
  low-\mstellar\ ($Z$) hosts, though this is more dependent on the
  split point used,
\item The $\alpha$ values in the split samples are consistent,
  differing only at $<1.5\sigma$,
\item SNe Ia in low-sSFR hosts show a smaller r.m.s. scatter about the
  best fits than those in high sSFR hosts; the r.m.s. scatter of SNe
  in high-\mstellar\ ($Z$) and low-\mstellar\ ($Z$) hosts are similar,
\end{itemize}
These results are not sensitive to the split points used to
characterise low-sSFR versus high-sSFR, or high-\mstellar\ ($Z$)
versus low-\mstellar\ ($Z$) hosts.  Fig.~\ref{fig:abbyssfrmass} shows
the joint confidence contours in the three combinations of the
nuisance variables for the sSFR and \mstellar\ split samples.

We also assess the significance of these results using a Monte-Carlo
permutation test of our data. We randomly draw $N_{\mathrm{low-sSFR}}$
(or $N_{\mathrm{high-M}}$ when considering the \mstellar\ split) SNe
Ia from the full SNLS sample (without replacement) to generate a fake
low-sSFR (high-\mstellar) sample with the same number of events as the
true low sSFR sample. The $N_{\mathrm{high-sSFR}}$
($N_{\mathrm{low-M}}$) remaining SNe are assigned to the fake high
sSFR (low-\mstellar) sample.  The nuisance variable fits are repeated
on both datasets, and the best-fitting $\alpha$, $\beta$ and \absm\
recorded. We repeat this procedure 25000 times, and compare the
distribution of derived nuisance variables from the permuted samples
to the values recovered for the real samples. We assess how frequently
the difference between the nuisance variables for the fake sample is
equal to or greater than the difference between the same parameters
measured on the real data.  The Monte Carlo approach can also assess
the significance of any differences seen in r.m.s. scatter between
different samples, otherwise difficult to assess statistically.

For samples split by sSFR, we find that the differences in $\alpha$
are found in randomly permuted samples 23\% (a $\sim1.2\sigma$
significance) of the time, for $\beta$ 2.6\% ($\sim2.2\sigma$), for
\absm\ 0.5\% ($\sim2.8\sigma$) and for the r.m.s. scatter 3.9\%
($\sim2.1\sigma$).  For samples split by \mstellar, the differences
are found in random samples 34\% ($\sim0.95\sigma$) for $\alpha$,
6.5\% ($\sim1.8\sigma$) for $\beta$, 0.005\% ($\sim3.5\sigma$) for
\absm, and 44\% ($\sim0.8\sigma$) for the r.m.s..  These results
support our main finding above: the most significant differences are
seen in \absm\ for all methods of splitting the host galaxies. Monte
Carlo tests show smaller significances in $\beta$ and r.m.s. scatter
for the sSFR split, and the differences in r.m.s. are not significant
for samples split by \mstellar. For the $\alpha$ parameters no
significant differences are seen. Monte Carlo tests on the $Z$-split
sample are similar to those for \mstellar.

\subsection{Effect of assumed cosmology}
\label{sec:effect-assum-cosm}

Our choice of cosmology could affect our results in two ways. The
first is the impact on the derived absolute host galaxy properties,
such as \mstellar\ or sSFR.  Changing the assumed $H_0$
(70\,km\,s$^{-1}$\,Mpc$^{-1}$) has only a superficial effect -- all
the host galaxy masses or other derived properties will change
relative to each other in the same consistent way.  Altering the other
cosmological parameters, such as \omatter, has a more subtle effect.
In a flat universe, a smaller \omatter\ will make the higher-redshift
hosts more massive compared to those at lower-redshift (effectively
they become more distant); a larger \omatter\ will have the opposite
effect. Within the errors to which cosmological parameters are
currently measured, this is a small effect -- a stellar mass derived
with $\omatter=0.30$ instead of $\omatter=0.256$ changes by $<$0.04 dex
at $z=0.5$ and $<$0.07 dex at $z=1$. We have run our fits using this
alternative model and find essentially identical results.

The cosmological model could also impact the SN properties. If the
redshift distributions of the host galaxies either side of the split
point in mass or metallicity are different, then any systematic trend
in SN brightnesses with redshift may increase (or decrease) the
significance of the differences in the nuisance parameters. For
example, an incorrect photometric zeropoint in one of the SNLS filters
would introduce a difference in SN Ia brightness which is a smooth
function of redshift, and if the mix of hosts also changes with
redshift, this could introduce a corresponding offset in the
magnitudes of SNe Ia in those hosts.

\begin{figure*}
\centering
\includegraphics[width=0.495\textwidth]{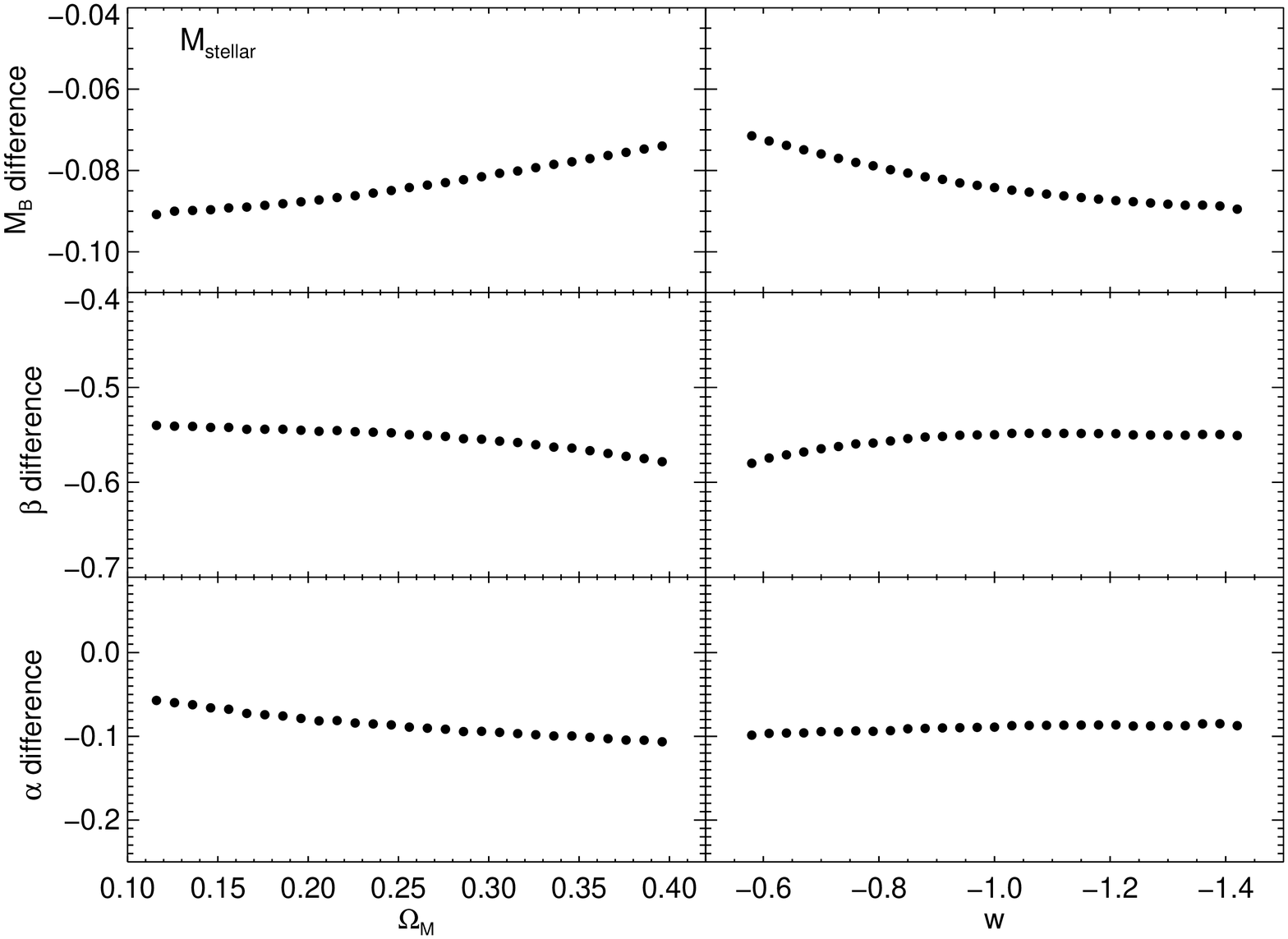}
\includegraphics[width=0.495\textwidth]{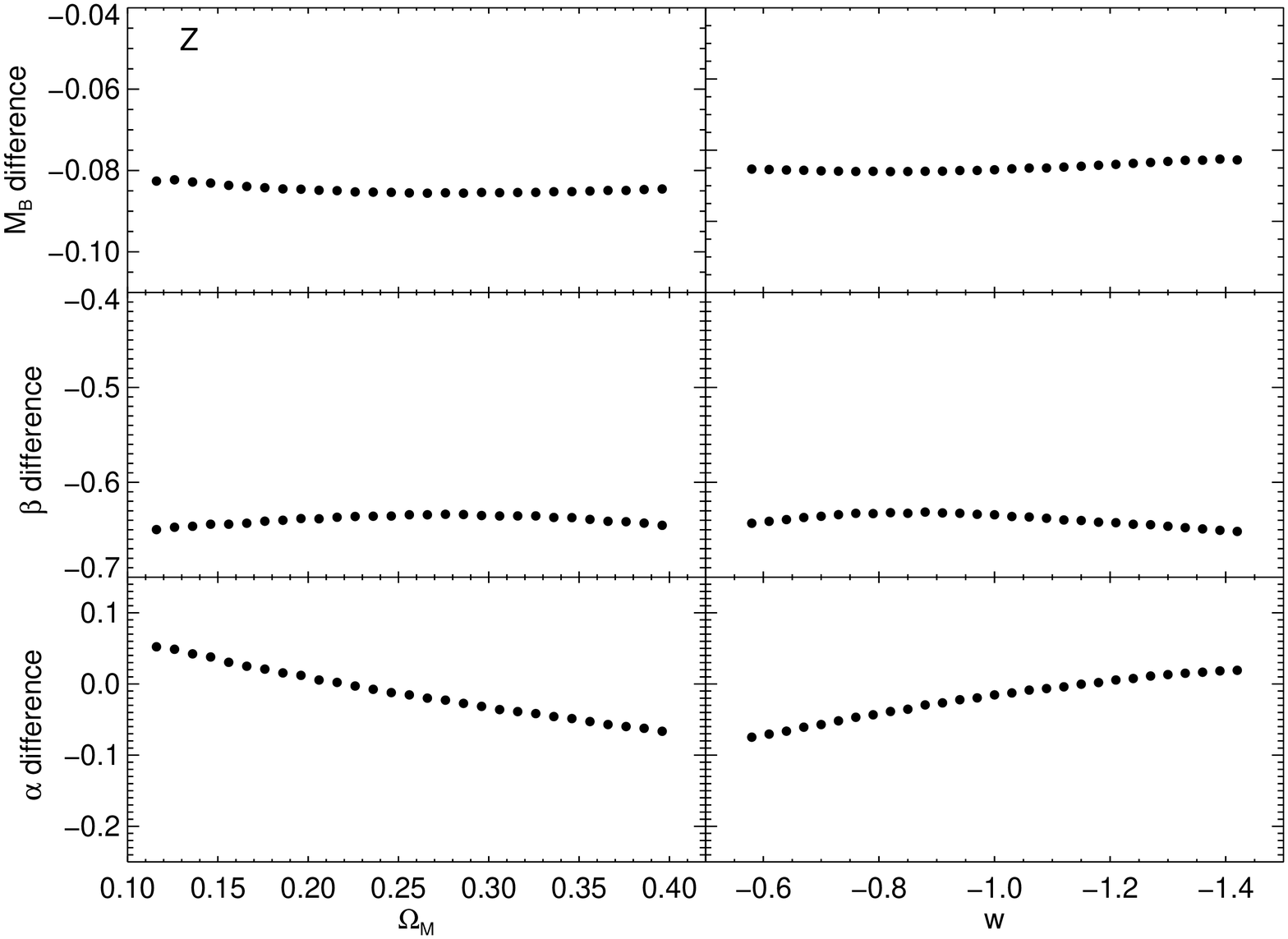}
\caption{The effect of the assumed cosmology on the differences in
  nuisance parameters derived from SNe Ia located in hosts either side
  of the default split points. The left-hand figure shows SN Ia hosts
  split by \mstellar, the right-hand figure by $Z$. Either \omatter\
  or $w$ is altered, with the other parameters held fixed, and the
  difference in nuisance parameters derived from SNe Ia in
  low-\mstellar\ and high-\mstellar\ (or low-$Z$ and high-$Z$) hosts
  plotted.\label{fig:cosmoeffect}}
\end{figure*}

We test the effect of the assumed cosmology in
Fig.~\ref{fig:cosmoeffect}. We start with our default cosmology
($\omatter=0.256$, $w=-1$), and vary \omatter\ (by $\pm$0.15) and $w$
(by $\pm$0.4) recording the difference in derived nuisance parameters.
As might be expected only small variations in the differences in the
derived nuisance parameters are seen. The largest variation in the
difference in \absm\ is with \mstellar\ ($\sim\pm0.01$), for $\beta$
with \mstellar\ ($\sim\pm0.05$), and for $\alpha$ with $Z$
($\sim\pm0.05$). Thus the assumption of the cosmological model does
not drive our results.

\subsection{SNLS selection effects}
\label{sec:selection-effects}

A final consideration is the possibility of a selection effect which
operates to either select against fainter SNe Ia (after correction) in
massive or low-sSFR galaxies, or brighter SNe Ia in low-\mstellar\ or
high-sSFR galaxies. A mechanism for the latter is difficult to
conceive, but for the former possible biases can be imagined. SNe Ia
in high-\mstellar\ galaxies will be \textit{intrinsically} fainter
(i.e., lower-stretch) and will lie against a brighter host background,
decreasing their contrast over their host galaxies. If this operates
near the SNLS spectroscopic limit, then we may be biased to
preferentially observe the brighter sub-sample of this population in
massive galaxies, i.e. those that at fixed light-curve shape are
over-luminous (following the \sigint\ distribution).

\begin{figure}
\centering
\includegraphics[width=0.495\textwidth]{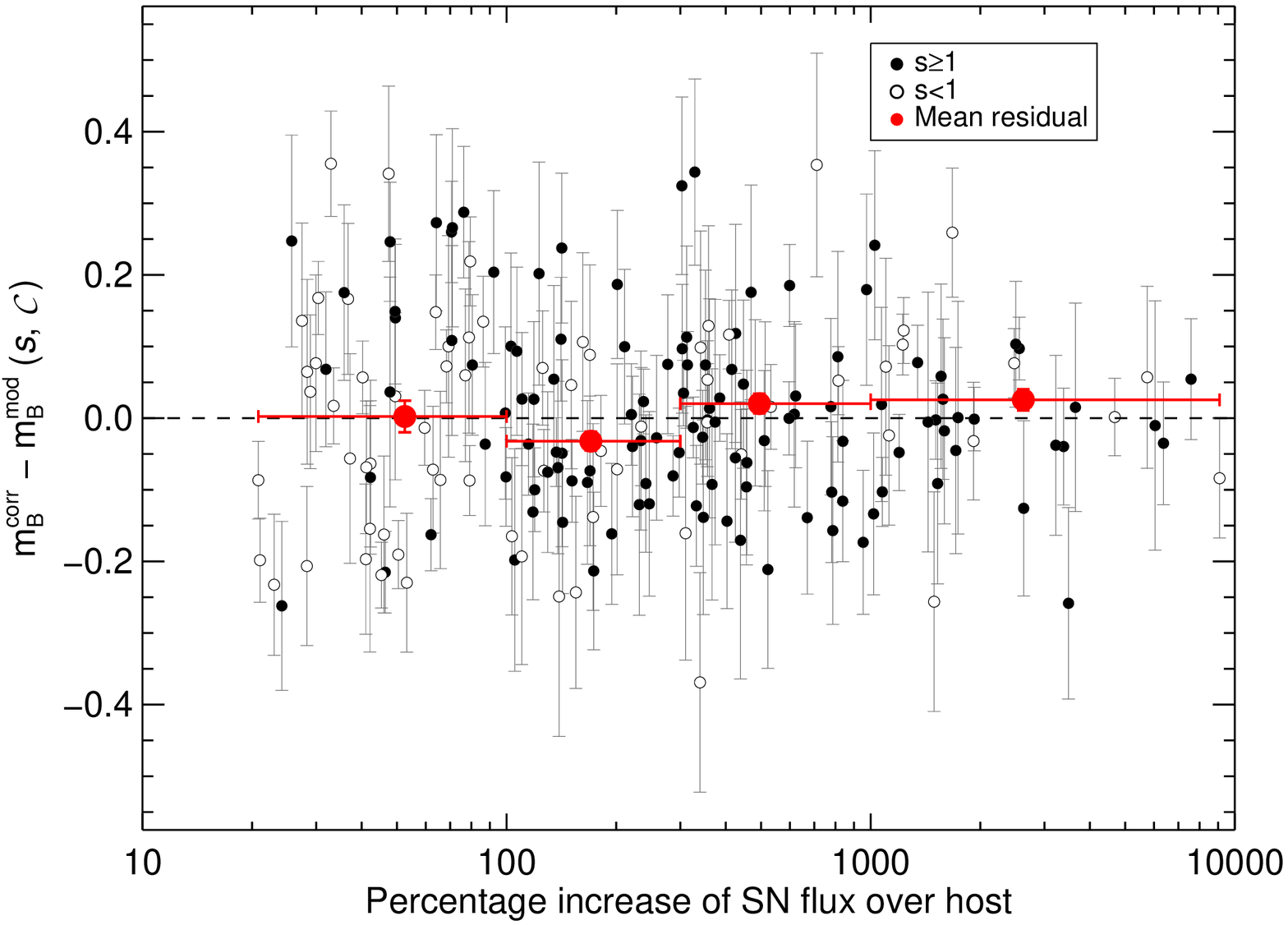}
\caption{Residuals from the best-fitting cosmology for the SNLS SNe
    as a function of the percentage increase of the events over their
    host galaxies at the time of spectroscopic observation. SNe Ia
    with $s<1$ are shown as open circles, and those with $s\geq1$ as
    filled circles. Brighter SNe Ia after correction have negative
    residuals. The red circles are the mean residuals in bins of
    percentage increase, drawn at the mean value of the percentage
    increase in each bin. The error bars on these points represent the
    bin width in percentage increase, and the weighted error on the
    mean for the residual axis.\label{fig:perinc}}
\end{figure}

We have partially mitigated against this selection effect by
restricting our analysis to those SNe Ia lying at $z<0.85$, away from
the redshift limit of SNLS. The total Malmquist bias (including
spectroscopic selection) on our SNe below this redshift is
$<0.015$\,mag \citep{2010perrettrta}, compared with the size of the
magnitude difference in our results of $\sim$0.1\,mag. We have also
tested for the existence of this observational selection effect by
examining the SN Ia residuals versus the percentage increase of the SN
flux over its host. Below a percentage increase of 100\% (i.e., the SN
is as bright as the host measured through a small aperture),
identification becomes more difficult
\citep[e.g.][]{2005ApJ...634.1190H}.  Our key diagnostic would be to
see brighter SNe (after correction) when the percentage increase is
$<$100\%, and fainter SNe at percentage increases $>$100\%, if this
selection effect were serious. We show these data in
Figure~\ref{fig:perinc}. No effect is present in our data; only weak
trends are present with the percentage increase as expected given the
weak correlation with \mstellar.

As a final test, we have also examined the Hubble residuals of events
which were not followed spectroscopically by SNLS, but which were
subsequently located in a offline search of the same data
\citep{2009A&A...499..653B} and are photometrically similar to SNe Ia
(Bazin et al., in prep.). The number of additional small
percentage-increase objects at $z<0.85$ is only 24 ($\sim$10\% of our
total sample), and have the same host-dependent trends as those
presented in this paper. We therefore conclude that this potential
selection effect cannot drive our results.

\section{Discussion}
\label{sec:discussion}

\subsection{Comparison to other results}
\label{sec:comp-other-results}

The results of $\S$~\ref{sec:lum-depend-trends} can be compared to
several studies from the literature. The result that high-\mstellar\
galaxies host the brightest SNe Ia after correction was also found by
\citet{2009arXiv0912.0929K} at a $\simeq$2.5$\sigma$ significance
using a similar low-redshift SN Ia sample to that used here.

Using galaxy morphologies for the same low-redshift sample,
\citet{2009ApJ...700.1097H} found weak trends with host galaxy type:
SNe Ia in Scd/Sd/Irr galaxies are fainter than those in elliptical
galaxies after correction (2$\sigma$ difference), and have a smaller
r.m.s. scatter and lower reddening. Assuming that sSFR closely tracks
galaxy morphology, we find a similar effect in that fainter SNe are
located in high sSFR galaxies, but we do not reproduce their result
that SNe in spiral galaxies have a smaller scatter on the Hubble
diagram; in fact our SNLS data suggest the opposite, and at
low-redshift we see no significant difference in r.m.s. scatter. We
also see no evidence that SNe in spirals have less reddening; SNe in
spirals appear redder than those in ellipticals.

Turning to metallicities, \citet{2005ApJ...634..210G} used spiral
galaxy emission line measurements to directly measure
$\log(\mathrm{O}/\mathrm{H})$ for 16 local host galaxies containing
SNe for which a Hubble residual was available, but found no
significant trend -- their results indicated that brighter SNe after
correction may lie in more metal rich galaxies, but only at 90\%
confidence.  \citet{2008ApJ...685..752G} used another small sample (17
host galaxies) of local SNe in E/S0 galaxies, using
$[\mathrm{M}/\mathrm{H}]$ spectral measures to show that brighter SNe
Ia (again, after correction) were hosted by more metal rich systems
with $\simeq$98\% confidence.

\citet{2009ApJ...691..661H} previously used 55 SNLS SNe Ia and
techniques similar to those in this paper, but found a correlation
between Hubble residual and inferred host metallicity (or \mstellar)
at only $\simeq1.3\sigma$.  These 55 events are also included in this
paper, but note that the Hubble residuals of
\citet{2009ApJ...691..661H} were based on photometry, calibration and
light-curve fitting tools used in \citet{2006A&A...447...31A} -- these
have all since improved \citep[see discussions
in][]{2007A&A...466...11G,2008ApJ...681..482C,2009A&A...506..999R,2009guylcs}.
The slope of Hubble residual with inferred
$\log(\mathrm{O}/\mathrm{H})$ as measured by
\citet{2009ApJ...691..661H} is $-0.10\pm0.07$; fitting the binned
points in the SNLS--metallicity plot in Fig.~\ref{fig:resid_ssfr}
gives a slope of $-0.18\pm0.06$, consistent with
\citet{2009ApJ...691..661H} at $\sim$1.5$\sigma$.

In summary, some similar trends to those presented in this paper have
been found in the past, but all at quite low significance. This is the
first dataset where a dependence of corrected SN Ia luminosities on
host properties has been definitively detected at $>$3$\sigma$
confidence.

\subsection{Colour residuals}
\label{sec:-colo-resid}

\begin{figure*}
\centering
\includegraphics[width=0.495\textwidth]{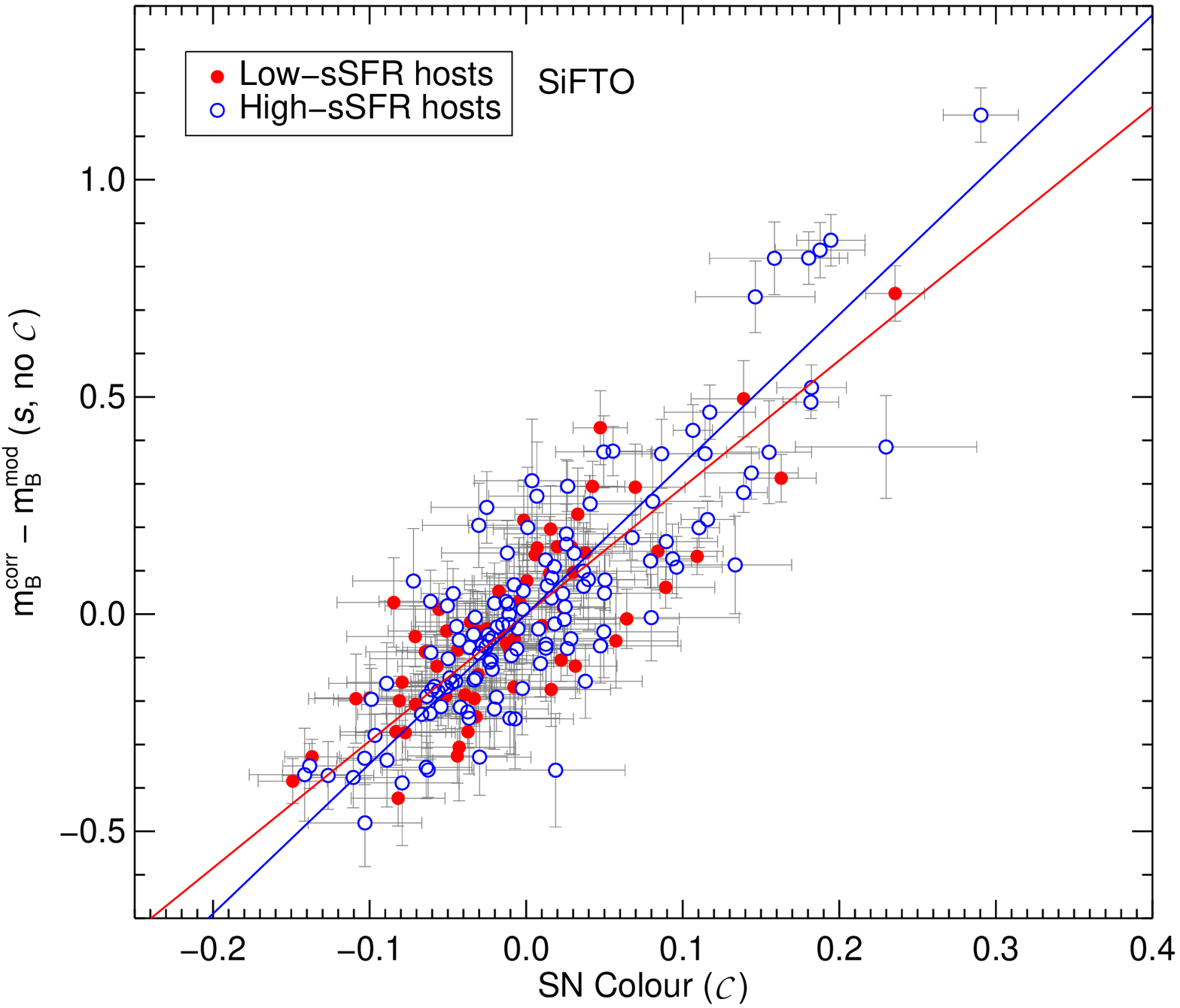}
\includegraphics[width=0.495\textwidth]{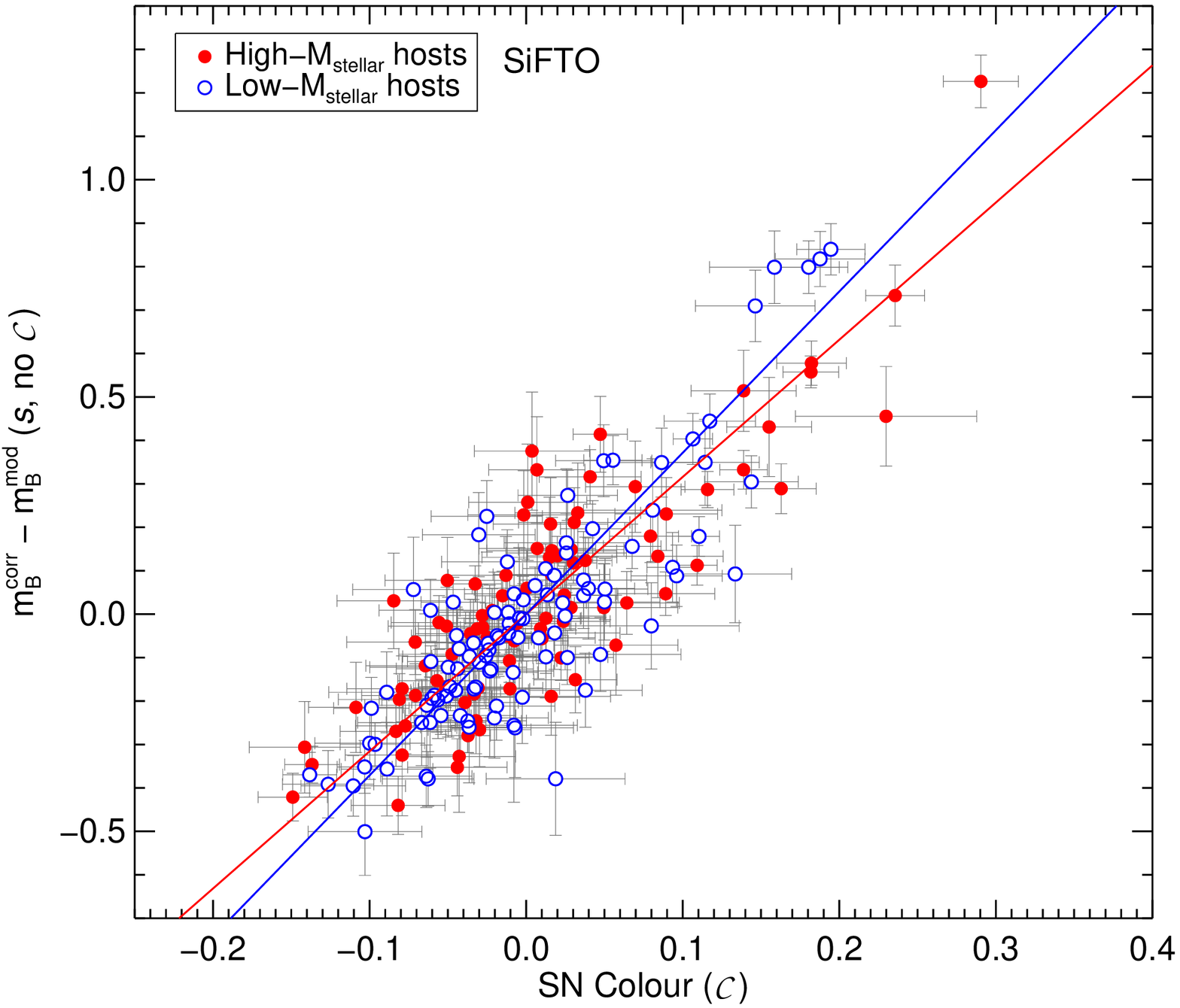}
\caption{Hubble diagram residuals as a function of \col\ for SNLS
  SNe~Ia before the colour--luminosity relation is applied, split by
  sSFR (left) and \mstellar\ (right), with the cosmological model
  fixed and the nuisance variables $\alpha$, $beta$, and \absm\
  allowed to vary according to the type of host.  The over-plotted
  lines show the best fitting relations (a slope of $\beta$). Note
  that the exact values of the residuals vary between the sSFR and
  \mstellar\ splits as the nuisance variables are allowed to change.
  As each set of hosts is allowed to have different values of \absm,
  the differences in SN Ia luminosity between different host types is
  not present.\label{fig:colresid}}
\end{figure*}

The Hubble residuals as a function of colour for SNLS SNe Ia
segregated by the sSFR and \mstellar\ of their hosts are shown in
Fig.~\ref{fig:colresid}, with the best-fitting relations
over-plotted. The (few) objects with red colours could influence the
best-fitting $\beta$ disproportionately. We repeat the nuisance
variable fits, each time reducing the maximum allowed colour by 0.05
and re-examining the fit $\beta$ (Table~\ref{tab:redcoltest}). Clearly
the errors in $\beta$ increase as the redder SNe are excluded due to
the reduced baseline in colour.

\begin{table*}
\centering
\begin{tabular}{ccccccccccc}
\hline
Reddest \col\ & \multicolumn{5}{c}{Low sSFR hosts}&\multicolumn{5}{c}{High sSFR hosts} \\
permitted     & $N_{\mathrm{SN}}$ & $\alpha$ & $\beta$ & \absm & r.m.s. & $N_{\mathrm{SN}}$ & $\alpha$ & $\beta$ & \absm & r.m.s.\\
\hline
0.30 &  65 & 1.377$\pm$0.158 & 2.730$\pm$0.228 & -19.208$\pm$0.016 & 0.120 & 131 & 1.573$\pm$0.150 & 3.444$\pm$0.137 & -19.149$\pm$0.014 & 0.147\\
0.25 &  65 & 1.377$\pm$0.158 & 2.730$\pm$0.228 & -19.208$\pm$0.016 & 0.120 & 130 & 1.521$\pm$0.150 & 3.361$\pm$0.142 & -19.152$\pm$0.014 & 0.145\\
0.20 &  64 & 1.373$\pm$0.158 & 2.730$\pm$0.228 & -19.208$\pm$0.016 & 0.120 & 129 & 1.549$\pm$0.151 & 3.443$\pm$0.152 & -19.149$\pm$0.014 & 0.143\\
0.15 &  63 & 1.435$\pm$0.164 & 2.947$\pm$0.264 & -19.202$\pm$0.016 & 0.123 & 122 & 1.614$\pm$0.145 & 3.214$\pm$0.183 & -19.149$\pm$0.013 & 0.137\\
0.10 &  61 & 1.548$\pm$0.170 & 3.346$\pm$0.328 & -19.190$\pm$0.017 & 0.129 & 113 & 1.600$\pm$0.148 & 3.617$\pm$0.257 & -19.136$\pm$0.014 & 0.140\\
\hline
\end{tabular}
\caption{As Table~\ref{tab:alphabeta_bysfr_sifto}, but testing the effect of the reddest SNe Ia on the derived nuisance variables. Similar results are found for SALT2 fits.\label{tab:redcoltest}}
\end{table*}

The difference between the $\beta$ in low and high sSFR galaxies
becomes less significant as the reddest allowed SN Ia colour is
reduced -- most of the difference seems caused by the reddest SNe Ia.
Though the colour variation in SNe Ia is not well understood, high
sSFR galaxies are likely to show a larger range in dust content than
low sSFR galaxies, and hence display a larger scatter and/or steeper
colour--luminosity relations for redder SNe Ia.  Removing these redder
events is likely to minimise the effect of any difference in $\beta$
between host galaxy types. It may also be the case that the
colour--luminosity relation of the reddest SNe is dominated by the
effects of dust rather than intrinsic variation (in which case
$\beta\sim R_B$), and hence redder SNe favour larger values of
$\beta$.  We note that \citet{2010AJ....139..120F} actually find an
decrease in $R_V$ ($R_V=R_B-1$) in a sample of low-redshift SNe Ia
when including very red SNe Ia, though these events with $(B-V)\sim1$
are much redder than any SNLS SN considered in this paper. They
speculate that this difference in $R_V$ in the reddest SNe Ia could be
a result of circumstellar dust \citep[see
also][]{2005ApJ...635L..33W}.

With smaller maximum allowed colours, the r.m.s. scatter of the fits
of SNe Ia in each type are also more consistent. Under the hypothesis
that the reddest SNe Ia are the most affected by dust this might be
expected.  SN Ia host galaxies are likely to display a range in
effective $R_V$ -- the Galactic average is 2.99$\pm$0.27
\citep[e.g.][]{2007ApJ...663..320F}, and there is no reason to expect
the range present in SN Ia hosts to be smaller. For an $E(B-V)$ of
0.1, this would generate an additional scatter of 0.06\,mag in SN peak
luminosities compared to a SN Ia with a colour excess of zero.

\subsection{SN stretch or host dependence?}
\label{sec:sn-stretch-or}

Given the dependence of SN Ia stretch on host galaxy parameters such
as \mstellar\ and sSFR (Fig.~\ref{fig:sc_ssfrmass}), the trends of SN
Ia brightness with the same parameters (Fig.~\ref{fig:resid_ssfr})
could simply arise from an incomplete stretch--luminosity correction,
rather than an intrinsic dependence on any third variable beyond
stretch and colour.  For example, if the stretch--luminosity relation
were more complex than a simple linear trend but only a linear
variable were fit, this could manifest as a dependence of corrected SN
Ia luminosity on host galaxy parameters.

We investigate this in two ways. We first examine the luminosity
trends as a function of host galaxy parameters segregated by stretch.
The SNe Ia plotted in Fig.~\ref{fig:sc_ssfrmass} are coded according
to the stretch of the event, and the residuals of both $s<1$ and
$s\geq1$ SNLS SNe Ia are given in Table~\ref{tab:snlsresids_byssfr}
(Table~\ref{tab:snlsresids_byssfr_salt} has the SALT2 equivalent).
Generally, the same trends with \mstellar, $Z$ and sSFR are seen for
high-$s$ and low-$s$ SNe Ia, though clearly the significance is
smaller than for the entire population given the number of events is
reduced.

\begin{figure}
\centering
\includegraphics[width=0.495\textwidth]{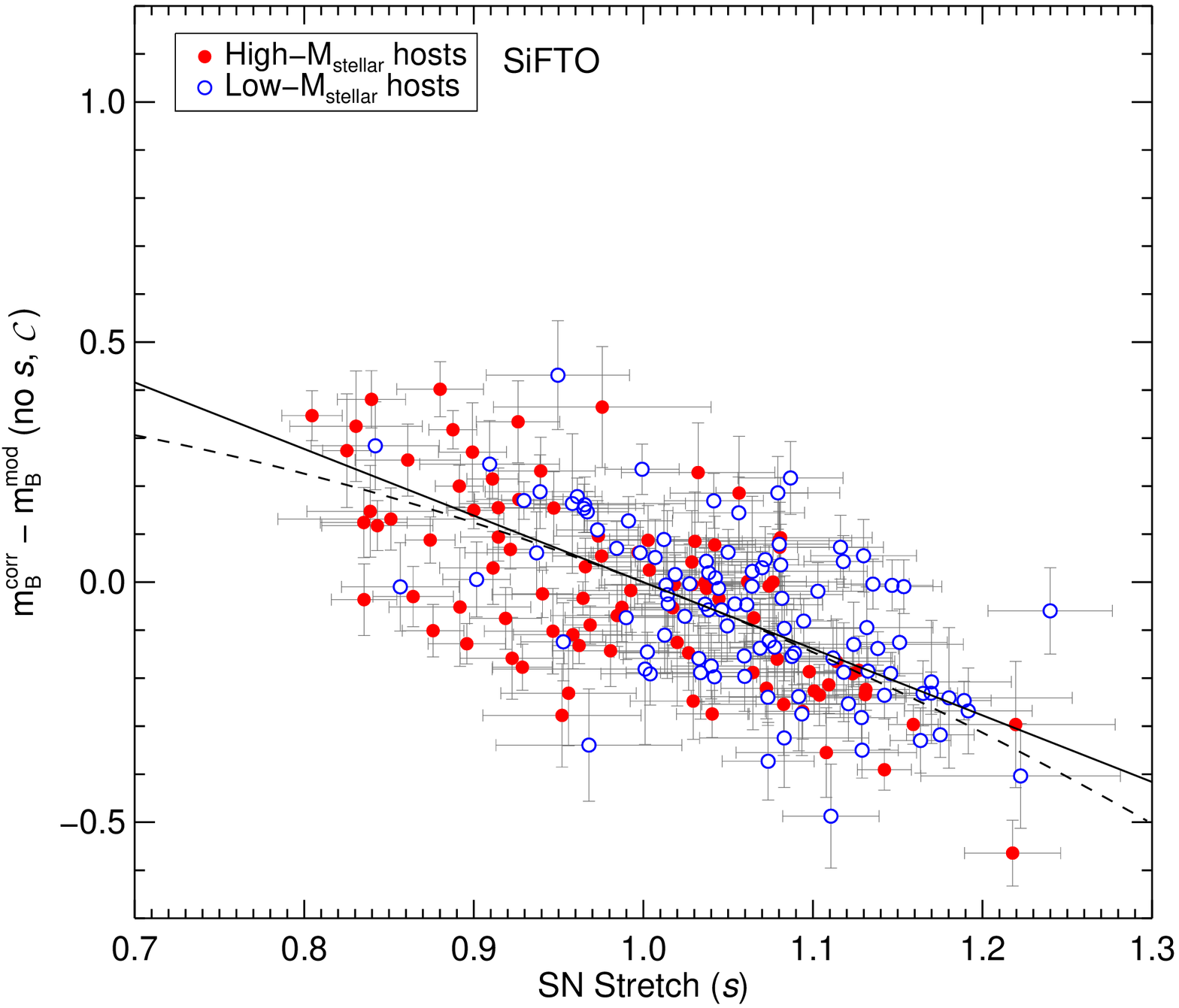}
\caption{Hubble diagram residuals as a function of $s$ for SNe~Ia
  split by \mstellar\ before the stretch--luminosity relation is
  applied. The over-plotted lines show the best fitting linear (solid)
  and quadratic (dashed) stretch--luminosity relations
  (Table~\ref{tab:massfits}). Only a single \absm\ is used, and so the
  SNe in high-\mstellar\ and low-\mstellar\ hosts are slightly
  offset.\label{fig:stretchresid}}
\end{figure}

We also experiment with a quadratic stretch--luminosity term in an
attempt to remove the host dependence. We add an additional term to
eqn.~(\ref{eq:mbcorr}) of the form $\alpha_2(s-1)^2$, shown in
Fig.~\ref{fig:stretchresid}. The improvement in the quality of the
fits with this extra stretch term is not significant
(Table~\ref{tab:massfits}) and does not remove the host-dependent SN
Ia luminosity trends. These tests suggest that the trends of SN Ia
luminosity with host parameters are largely independent of SN Ia
stretch.

\subsection{Cosmological implications}
\label{sec:cosm-impl}

Differences in mean SN Ia properties when split by host galaxy
properties, which are not removed by corrections currently employed in
cosmological analyses, could clearly lead to systematic errors in
cosmological analyses. Observationally, SNe Ia in massive galaxies
appear brighter than those in less massive galaxies -- a similar
effect is seen when considering sSFR, with SNe Ia in low sSFR galaxies
brighter than those in high sSFR galaxies. These differences are
significant at $>$3$\sigma$.

\begin{figure*}
\centering
\includegraphics[width=0.495\textwidth]{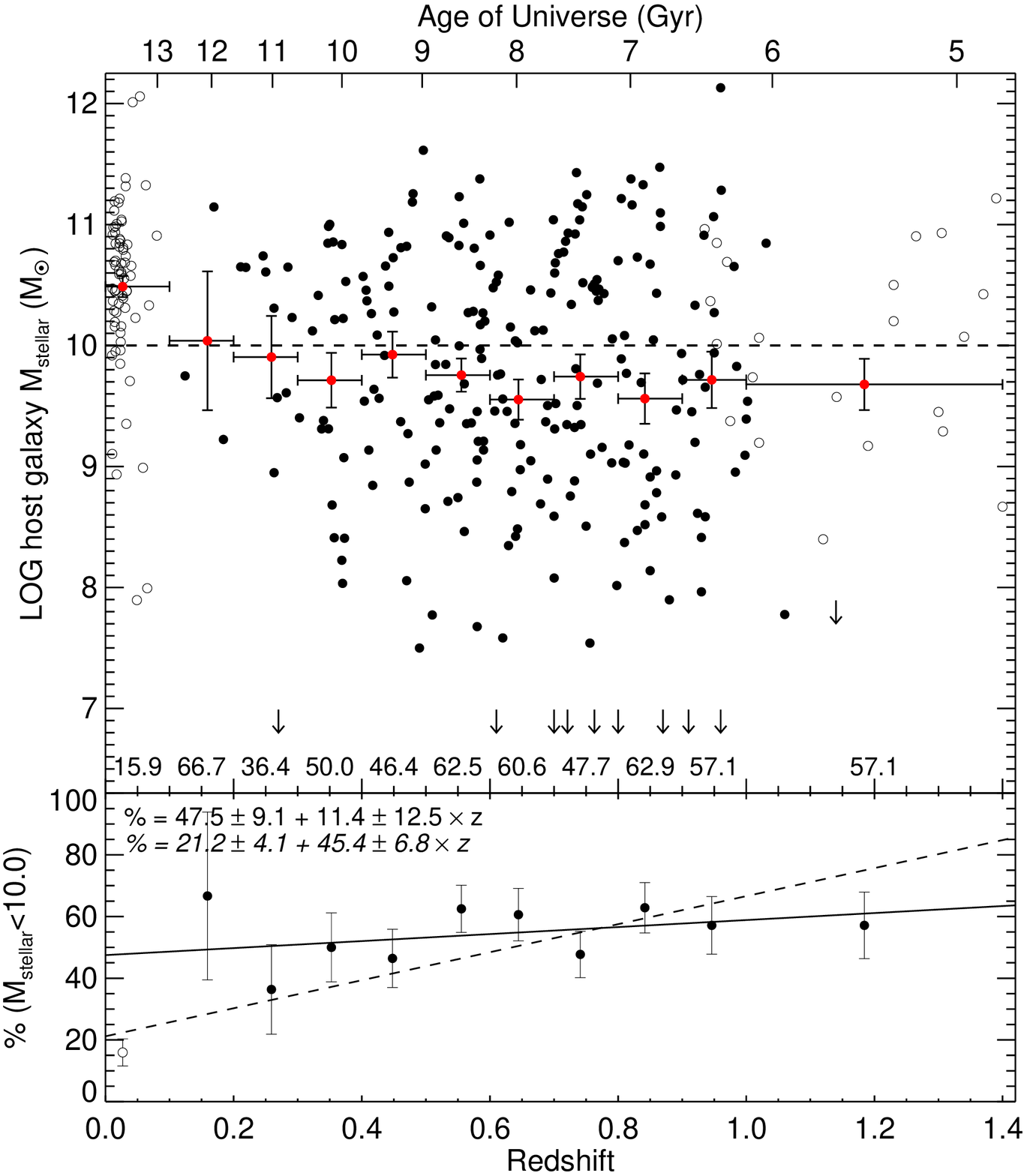}
\includegraphics[width=0.495\textwidth]{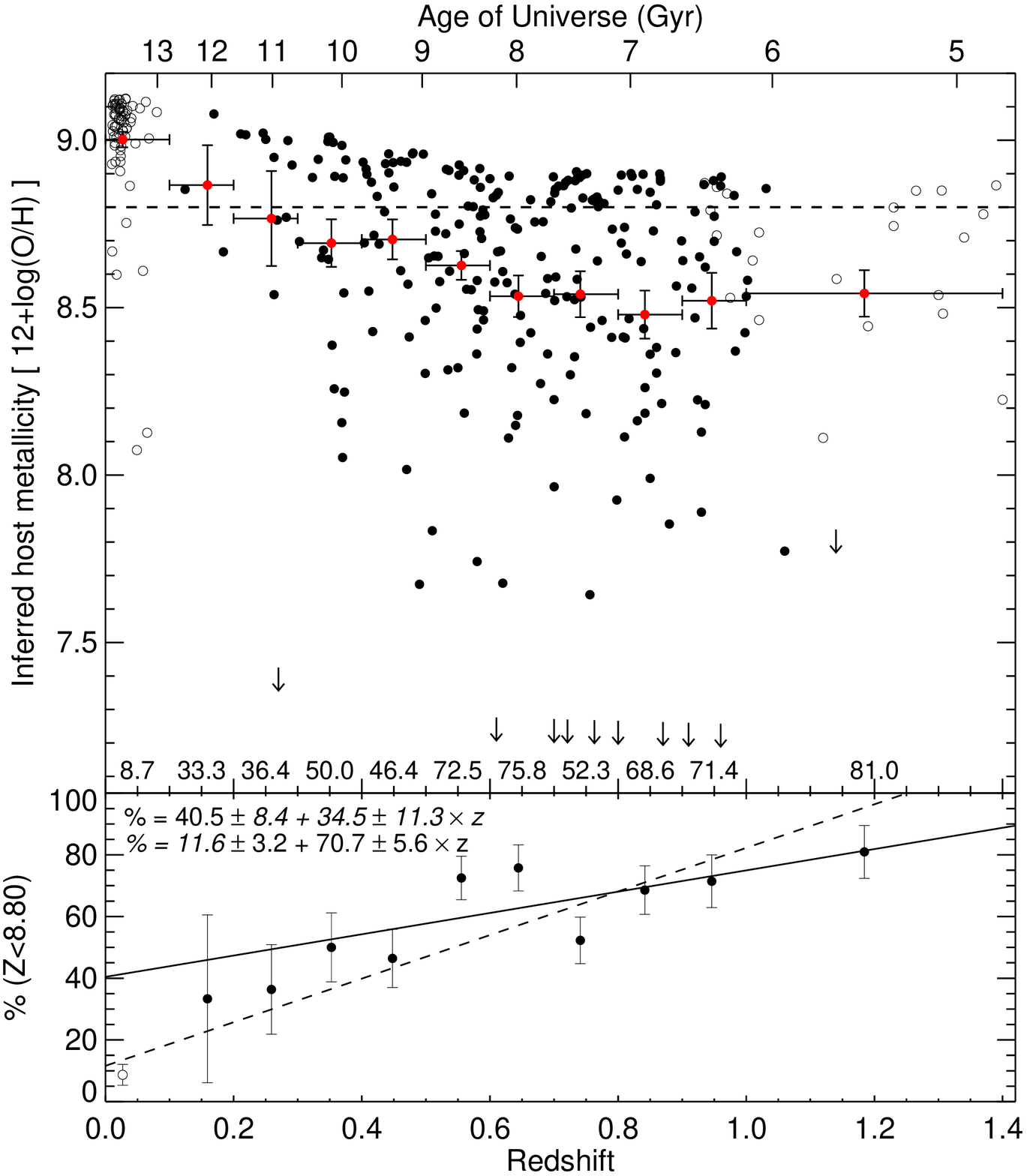}
\caption{The change with redshift in SN Ia host galaxy \mstellar\
  (left) and inferred $Z$ (right). The mean values in bins of redshift
  are shown as red circles, drawn at the mean redshift in each bin,
  and with error bars representing the bin width for the redshift
  axis, and the error on the mean in the \mstellar\ and $Z$ axis. The
  horizontal dashed lines show the split points used in
  $\S$~\ref{sec:lum-depend-trends}. Open circles refer to the low
  redshift or R07 samples, and filled circles the SNLS sample.  Note
  that the differences between the low-redshift and SNLS hosts is
  unlikely to be a real physical effect, and is generated by selection
  effects in the low-redshift sample. The lower panels show simple
  linear fits to the fraction of SN Ia hosts with
  $\log(\mstellar)10$\msun (or $Z<8.8$), including (dashed line) and
  excluding (solid line) the low-redshift data (shown as an open
  circle).  The fit parameters are shown; the numbers in italics
  include the low-redshift data.\label{fig:redshiftmassmetal}}
\end{figure*}

Evidence for two populations of SNe Ia with different photometric
properties is not by itself alarming, as the nuisance variables in any
global fit will average to values appropriate for the combination of
SNe Ia. However, any change in the mix of SNe Ia with redshift could
introduce a serious effect. We measure the change in SN Ia host
\mstellar\ or $Z$ as a function of redshift
(Fig.~\ref{fig:redshiftmassmetal}) using the SNLS dataset unrestricted
in redshift range, and the low redshift and R07 data. We measure the
fraction of host galaxies with \mstellar\ or $Z$ less than the split
points defined in earlier sections in each redshift bin. We then make
a simple linear fit to these values as a function of redshift i.e.,
$a+bz$, where $a$ and $b$ are the fit coefficients. To guard against
selection effects, we perform the fits with and without the
low-redshift data where the selection effects are quite different to
those in the SNLS data.  In the case of metallicity, the strength of
any redshift trend is driven by the form of the \mstellar--metallicity
relationship assumed -- our default relation that evolves with
redshift leads to a correspondingly larger redshift trend than a
relation that is assumed constant with redshift (and also leads to a
decreasing upper metallicity limit for any given redshift, as seen in
Fig.~\ref{fig:redshiftmassmetal}). If a relation with no redshift
evolution is used, the metallicity--redshift plot becomes essentially
the same as the \mstellar--redshift plot.

As expected from a consideration of popular galaxy formation models,
the host galaxies at higher redshift contain, on average, less stellar
mass and consequently are likely to be of lower metallicity.  The
strength of the evolution is most significant with metallicity in the
SNLS+R07 data, but is substantially strengthened in \mstellar\ if the
low-redshift host data are also included. However, including the
low-redshift data is likely to overestimate the strength of the real
physical evolution, as the bulk of the low-redshift is strongly biased
against host galaxies of low stellar mass. It would be appropriate to
include the low-redshift data if the amount of evolution likely in
current cosmological samples was required to be estimated.

Given the expectation and evidence of evolution, either from selection
effects or an underlying physical effect, we investigate methods for
accounting for it in cosmological analyses, and the effect that it
would have if left uncorrected.

\subsubsection{Additional nuisance variables}
\label{sec:addit-nuis-param}

We consider two ways to include additional parameters in the fits. The
first and most obvious is to include a further linear $\gamma\times x$
term to eqn.~(\ref{eq:mbcorr}), where $x$ is a variable such as
\mstellar, $Z$ or sSFR measured from the SN Ia host galaxies. The
second is motivated by the idea of two populations of SNe Ia rather
than a continuum in properties, and instead introduces two independent
values for \absm\ in eqn.~(\ref{eq:mtheory}) for SNe Ia located in
different types of host galaxy. Explicitly, SNe Ia located in galaxies
with $\log(\mstellar)\geq10.0$ (or $Z\geq8.8$, or $\log(\mathrm{sSFR})\le-9.7$)
are assigned \absmhm, and SNe Ia in galaxies with
$\log(\mstellar)<10.0$ (or $Z<8.8$, or $\log(\mathrm{sSFR})>-9.7$) are assigned
\absmlm. In principle, different values of $\alpha$ and $\beta$ could
also be fit for SNe Ia lying either side of the split point. However,
the results of $\S$~\ref{sec:test-univ-nuis} show that these nuisance
variables are more consistent than \absm, i.e. the \absm\ captures
most of the host galaxy dependent variation.

The results of these different fits with the three different host
variables are given in Table~\ref{tab:massfits} (again, the cosmology
is held fixed). We also hold \sigint\ fixed, so the addition of an
extra parameter into these fits leads to a significant reduction in
$\chi^2$, particularly for the \mstellar\ and related metallicity
fits. $F$-tests show that the addition of an extra parameter improves
the quality of the fits beyond that expected by the addition of the
extra parameter. For the two \absm\ fits, the significance is
3.9--4.4$\sigma$ for \mstellar\ and 4.1--4.4$\sigma$ for $Z$. The
linear fit, though still improving the $\chi^2$, generally performs
more poorly: 3.3--3.5$\sigma$ for \mstellar\ and 2.6--3.0$\sigma$ for
$Z$. Fits for two \absm\ using \mstellar\ or $Z$ perform better than
those with sSFR; note that the linear fits in sSFR are difficult to
implement as a zero SFR cannot be represented in $\log(\mathrm{sSFR})$
and must be assigned a fixed nominal value.

\begin{table*}
\centering
\begin{tabular}{lccccccc}
\hline
Sample & $\alpha$ & $\beta$ & $\gamma$ or $\alpha_2$ & \absm\ or \absmhm & \absmlm & $\chi^2$ & DOF\\
\hline
\multicolumn{8}{l}{Base fits:} \\
                          SNLS & 1.384$\pm$0.103 & 3.348$\pm$0.124 & ... & -19.170$\pm$  0.010 & ... & 190.8 & 192\\
                  SNLS+low-$z$ & 1.353$\pm$0.083 & 3.127$\pm$0.100 & ... & -19.178$\pm$  0.008 & ... & 270.4 & 261\\
            SNLS, two $\alpha$ & 1.349$\pm$0.107 & 3.347$\pm$0.124 & 1.094$\pm$0.943 & -19.160$\pm$  0.011 & ... & 189.8 & 191\\
    SNLS+low-$z$, two $\alpha$ & 1.346$\pm$0.083 & 3.121$\pm$0.101 & 0.572$\pm$0.733 & -19.172$\pm$  0.008 & ... & 270.0 & 260\\
\hline
\multicolumn{8}{l}{\mstellar\ fits:} \\
        SNLS, $\gamma$ & 1.553$\pm$0.115 & 3.365$\pm$0.127 &  0.0423$\pm$ 0.0125 & -19.172$\pm$  0.010 & ... & 179.9 & 191\\
        SNLS, two \absm & 1.571$\pm$0.115 & 3.382$\pm$0.127 & ... & -19.209$\pm$  0.015 & -19.127$\pm$  0.015 & 176.0 & 191\\
 SNLS+low-$z$, $\gamma$ & 1.515$\pm$0.093 & 3.150$\pm$0.103 &  0.0412$\pm$ 0.0106 & -19.176$\pm$  0.009 & ... & 256.4 & 260\\
SNLS+low-$z$, two \absm & 1.508$\pm$0.091 & 3.178$\pm$0.103 & ... & -19.211$\pm$  0.011 & -19.132$\pm$  0.014 & 251.4 & 260\\
\hline
\multicolumn{8}{l}{Metallicity fits:} \\
         SNLS, $\gamma$ & 1.521$\pm$0.116 & 3.398$\pm$0.129 &  0.1124$\pm$ 0.0412 & -19.180$\pm$  0.011 & ... & 184.3 & 191\\
        SNLS, two \absm & 1.587$\pm$0.116 & 3.377$\pm$0.127 & ... & -19.214$\pm$  0.015 & -19.127$\pm$  0.015 & 175.2 & 191\\
 SNLS+low-$z$, $\gamma$ & 1.482$\pm$0.092 & 3.184$\pm$0.104 &  0.1107$\pm$ 0.0328 & -19.181$\pm$  0.009 & ... & 261.0 & 260\\
SNLS+low-$z$, two \absm & 1.520$\pm$0.092 & 3.172$\pm$0.103 & ... & -19.212$\pm$  0.011 & -19.132$\pm$  0.014 & 250.9 & 260\\
\hline
\multicolumn{8}{l}{sSFR fits:} \\
         SNLS, $\gamma$ & 1.580$\pm$0.115 & 3.254$\pm$0.127 & -0.0438$\pm$ 0.0118 & -19.176$\pm$  0.010 & ... & 178.4 & 191\\
        SNLS, two \absm & 1.525$\pm$0.113 & 3.295$\pm$0.125 & ... & -19.210$\pm$  0.017 & -19.144$\pm$  0.013 & 182.2 & 191\\
 SNLS+low-$z$, $\gamma$ & 1.541$\pm$0.094 & 3.078$\pm$0.102 & -0.0434$\pm$ 0.0101 & -19.181$\pm$  0.009 & ... & 253.7 & 260\\
SNLS+low-$z$, two \absm & 1.470$\pm$0.090 & 3.107$\pm$0.101 & ... & -19.211$\pm$  0.013 & -19.152$\pm$  0.012 & 259.6 & 260\\
\hline
\end{tabular}
\caption{Fits involving extra nuisance parameters based on host galaxy measurements, specifically \mstellar, $Z$ and sSFR. In all cases the cosmology is held fixed.\label{tab:massfits}}
\end{table*}

In the SNLS plus low-redshift fits with two \absm, \absmlm\ is more
poorly measured than \absmhm.  This can be traced to the biased nature
of the low-redshift sample. In SN Ia only fits, the absolute
magnitudes are heavily influenced by the low-redshift data where the
SN Ia brightness is less dependent on the cosmology, and as there are
far fewer SNe Ia in the low-redshift sample in the
low-\mstellar/low-$Z$ group, the statistical precision of \absmlm\ is
correspondingly reduced compared to \absmhm\ where there are abundant
low-redshift SNe Ia.

The use of two absolute magnitudes has other advantages over a linear
host parameter term. There is no theoretical reason to suppose that
any dependence on metallicity be linear; \citet{2003ApJ...590L..83T}
show that the putative effect will be strongly non-linear. From a
practical viewpoint, the use of two absolute magnitudes is
substantially less sensitive to systematics in host galaxy parameter
determination. The categorisation of low-\mstellar\ versus
high-\mstellar\ is not sensitive to systematic errors associated with
the choice of base SED library or assumptions about IMF or dust
extinction, nor to the statistical errors on the masses, particularly
at low-\mstellar\ where these can be considerable.

The other nuisance parameters are quite stable to the introduction of
a $\gamma$ or \absmlm\ term. $\beta$ is almost unchanged, and $\alpha$
tends to drift to larger values by about 0.15. This latter effect can
be understood when considering the stretch dependence on host galaxy
properties. Though the trends in SN Ia brightness are the same for
both low and high stretch SNe ($\S$~\ref{sec:sn-stretch-or}),
lower-$s$ SNe Ia are more prevalent in massive and low sSFR galaxies.
These SNe Ia prefer a brighter \absm\ than those in low-\mstellar\
hosts, making their Hubble residuals, when fitting for two \absm, more
positive (fainter). This in turn will lead to a systematic steepening
of the stretch--luminosity relation, which is reflected in the larger
values of $\alpha$ seen when using third parameters.

\subsubsection{Systematics in cosmology}
\label{sec:addit-syst-terms}

What is the size of the effect of this host galaxy dependence in
cosmological analyses? We analyse this by fitting the low-redshift,
SNLS and R07 data (314 SNe Ia) with two simple cosmological models
with and without the host galaxy term, and assessing the variation of
the fit parameters. For simplicity, we use only the statistical errors
of the SNe -- in a full cosmological analyses, the covariance matrix
between different SNe accounting for all systematic errors should be
used \citep{2009conleysys}; here we only aim to assess the size of the
effect. Our cosmological models are a flat, $\Lambda$CDM model
(fitting only for \omatter, i.e. assuming a constant $w=-1$), and one
in which $w$ is still constant but not required to be equal to $-1$
(fitting for \omatter\ and $w$). In each case the \sigint\ for each
SN sample is adjusted to give a reduced $\chi^2$ of 1.0.  We also show
the effect of adding external constraints into the fit using the
Baryon Acoustic Oscillation (BAO) constraints of
\citet{2005ApJ...633..560E}.  The fit results, expressed as a shift in
the cosmological parameters from their values without the host
parameter, can be found in Table~\ref{tab:cosmoeffect}. We use the
\texttt{cosfitter}
program\footnote{\texttt{http://qold.astro.utoronto.ca/conley/simple\_cosfitter/}},
written by one of us (AC) and adjusted to fit for two \scriptm, to
perform the fits. We test both \mstellar\ and gas-phase $Z$ as the
third variable on which to split the data.

\begin{table*}
\centering
\begin{tabular}{lccccccccc}
Third & $\Lambda$CDM & \multicolumn{4}{c}{$w$ model, SN only} & \multicolumn{4}{c}{$w$ model with BAO}\\
variable& $\delta$\omatter\ ($\sigma$) & $\delta$\omatter\ ($\sigma$) & $\delta w$ ($\sigma$) & \absmhm\ & \absmlm\ & $\delta$\omatter\ ($\sigma$) & $\delta w$ ($\sigma$) & \absmhm\ & \absmlm\ \\
\hline
 \mstellar & 0.012 (0.6) & 0.068 (0.8) & 0.173 (0.8) & -19.12$\pm$0.03 & -19.19$\pm$0.02 & 0.001 (0.1) & 0.040 (0.6) & -19.13$\pm$0.02 & -19.20$\pm$0.02\\
       $Z$ & 0.031 (1.6) & 0.066 (0.8) & 0.205 (1.0) & -19.10$\pm$0.03 & -19.19$\pm$0.02 & 0.006 (0.3) & 0.078 (1.2) & -19.11$\pm$0.02 & -19.20$\pm$0.02\\
\hline
\end{tabular}
\caption{Simple effect of a host galaxy parameter on cosmological constraints. The shift in the cosmological parameters are shown when a host parameter is added, together with the shift expressed as a fraction of the simple statistical error-bar. A full cosmological analysis is presented in \citet{2009conleysys} and \citet{2009sullivancosmo}.\label{tab:cosmoeffect}}
\end{table*}

\begin{figure}
\centering
\includegraphics[width=0.495\textwidth]{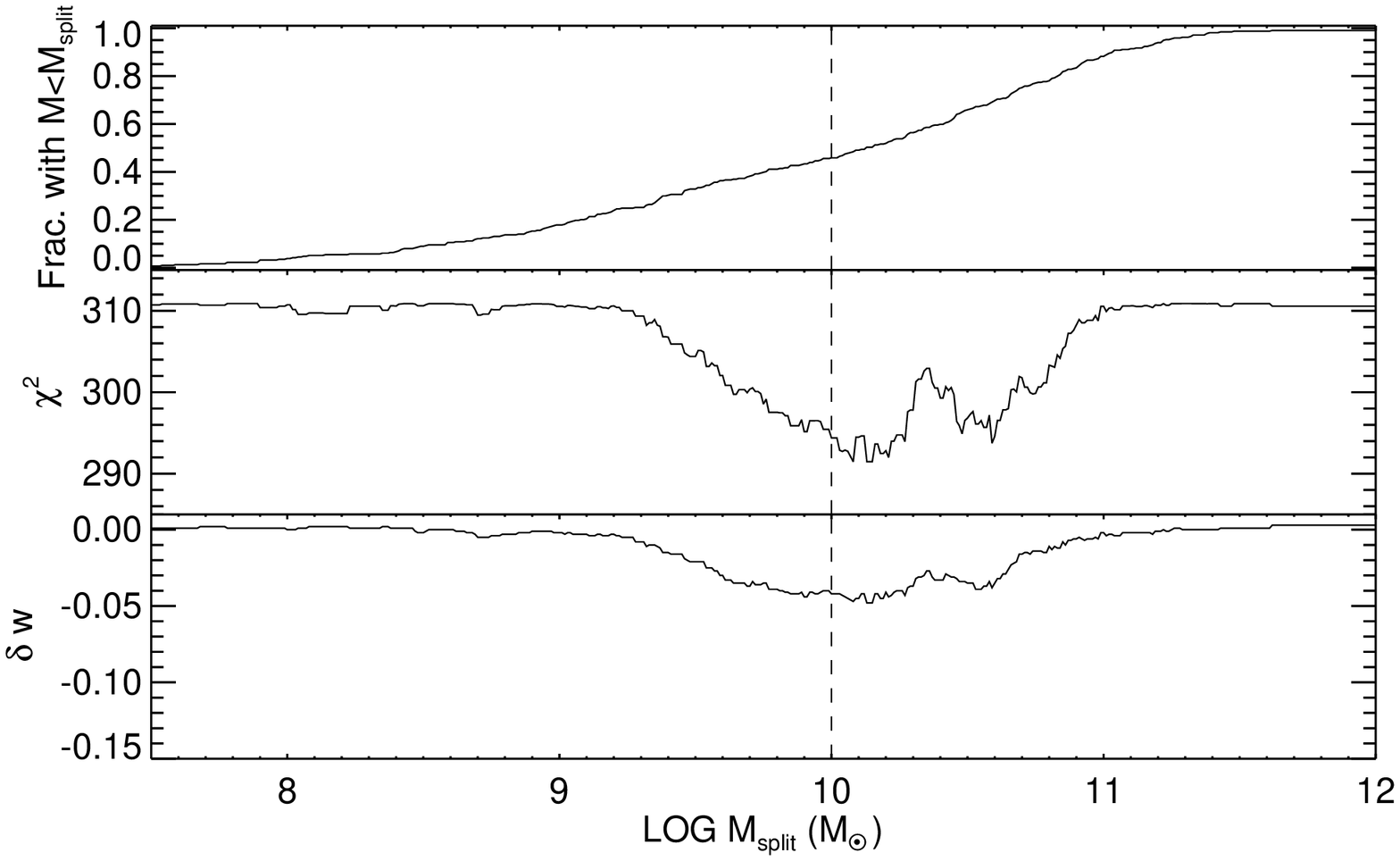}
\includegraphics[width=0.495\textwidth]{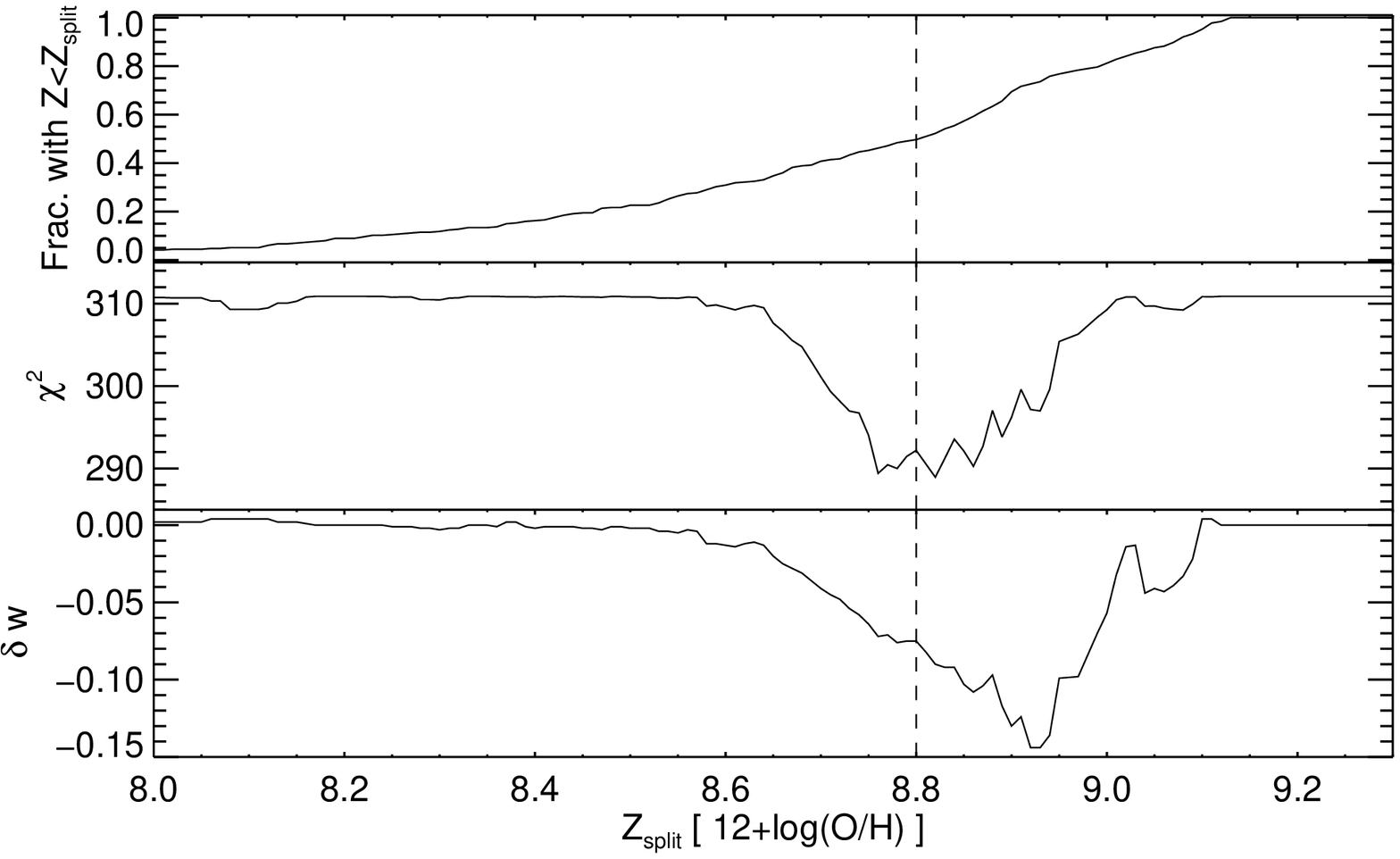}
\caption{The sensitivity of the cosmological results on the choice of
  split point used for the host galaxy variable. Cosmological fits are
  performed at each choice of split point, and the fraction of SNe Ia
  with the host parameter less than the split point (upper panels),
  the $\chi^2$ of the fit (middle panels), and the change in $w$
  (lower panels) are plotted as a function of the split point.  The
  upper plots show \mstellar, the lower ones $Z$. The default values
  are shown as vertical lines.  The $w$ measures include BAO
  constraints. When all SNe Ia are assigned to one of the groups as
  happens on the left and right edges of these plots, the cosmological
  results are identical.  \sigint\ is held fixed in these tests at the
  values required to produce a reduced $\chi^2$ of one without
  additional parameters -- the decrease in $\chi^2$ can be interpreted
  as a decrease in \sigint\ when including the host
  parameter.\label{fig:splitchoice}}
\end{figure}

The size of the effect -- the shift in the cosmological parameters
with a third variable -- is comparable to the simple statistical
precision (Table~\ref{tab:cosmoeffect}). These shifts are larger when
metallicity is the third variable ($>1\sigma$) instead of \mstellar\
($<1\sigma$). One source of systematic error is the choice of split
point in the host variable.  We investigate the dependence of the
cosmological results on this choice in Fig.~\ref{fig:splitchoice}. We
show the effect on $w$ (including BAO constraints) and the $\chi^2$
as the split point is varied. The median error in mass is $\simeq0.14$
dex ($\simeq0.03$ once converted to gas phase $Z$), and the relative
change in $w$ as the split points are varied over this range is
small.  One approach would be to iterate on the split point in
cosmological fits in order to find the model with the smallest
$\chi^2$.

As a final point, we note that the difference in $\beta$ between
low-sSFR and high-sSFR host galaxies, and a change in the mix of the
two with redshift, will introduce a mild redshift evolution in the
mean value of $\beta$. Such an evolution has been claimed
\citep{2009ApJS..185...32K}, though is not present to any significant
extent in SNLS3 data \citep{2009guylcs}. Given the small evolution in
\mstellar\ or sSFR with redshift in the SNLS data, we estimate any
$\beta$ evolution would be $<$0.1 across our entire redshift range,
below the detectability of our current dataset.

\subsection{Comparison to theory}
\label{sec:comparison-theory}

\citet{2003ApJ...590L..83T} showed that the amount of $^{56}$Ni
generated in a SN Ia explosion will depend on the metallicity of the
progenitor star. More metal-rich stars will generate increased
$^{14}$N during hydrogen burning as C, N and O are converted to
$^{14}$N in the CNO cycle. During core helium burning, this $^{14}$N
is converted to neutron-rich $^{22}$Ne, which will favour the
production of stable, neutron-rich $^{58}$Ni at the expense of the
radioactive $^{56}$Ni that drives the luminosity of SNe Ia. Fainter
SNe Ia should therefore be produced in more metal-rich environments.

While at first sight this may appear to disagree with the results in
Figs.~\ref{fig:resid_mass} and \ref{fig:resid_metal} which show the
opposite trend, \citet{2009Natur.460..869K} show this metallicity
effect will also affect the SN light-curve duration -- SNe Ia in
metal-rich environments are both fainter \textit{and} have narrower
light curves (or smaller stretches).  Though this trend operates in
the same sense as the global stretch--luminosity relation, and is
therefore partially corrected by it, the correspondence is not exact
and a difference is expected -- for a fixed light-curve width, a
higher metallicity SN Ia will be brighter. Though the slope of the
metallicity-driven stretch--luminosity relation is expected to change
with metallicity, over the stretch range that our SNLS data sample
(80\% of our SNe have $0.9<s<1.2$, or $1.2>\Delta m_{15}>0.8$ using
the relation from \citet{2008ApJ...681..482C}),
\citet{2009Natur.460..869K} show that the metallicity should act with
a fairly constant gradient in stretch--luminosity space. The effect of
metallicity should therefore translate to a simple offset in the
stretch--luminosity relation derived from our data, rather than a
change in slope. This is equivalent to a change in \absm, with
brighter SNe in metal-rich environments. Thus this theoretical
expectation of progenitor metallicity on SN Ia luminosity appears
qualitatively consistent with our data. (Note that these models also
predict that SNe Ia in massive galaxies should have slightly smaller
stretches due to their high metallicities, all other variables being
equal.)

Clearly, other physical variables correlate with host properties such
as \mstellar\ and sSFR, including stellar age and dust content
\citep[see discussion in][]{2009arXiv0912.0929K}.  Dust appears less
likely to be responsible for the effects in the SNLS data. The
colour--luminosity relations of SNe Ia in low-\mstellar\ and
high-\mstellar\ hosts are consistent, as are the mean colours of the
SNe in those hosts, as well as the distribution of those colours.
There is also the expectation that SNe Ia will be affected by only
small amounts of extinction, particularly in flux limited surveys like
SNLS \citep{1998ApJ...502..177H,2004NewAR..48..567C}.

The role of the progenitor system stellar-age is substantially more
complex, and clearly cannot be ruled out in our data -- the most
massive hosts will on average contain stars of older ages. At the time
of writing, no clear prediction has been made as to the effect of
progenitor age on SN Ia luminosity. There is speculation that
progenitor age drives the global relationship between light-curve
shape and host galaxy type, where morphologically earlier-type
galaxies host lower-stretch SNe Ia. Of importance in this context is
the result that the \mstellar--SN-brightness dependency does not depend
on the stretch of the SN. Low-$s$ SNe in lower-\mstellar\ hosts have a
similar corrected brightness as high-$s$ SNe in low-\mstellar\ hosts,
both consistently fainter than their counterparts in high-\mstellar\
hosts. Under the assumption that age is the major driver of the
stretch--luminosity relationship, this would suggest that it is
metallicity that is responsible for the SN Ia \mstellar--luminosity
dependence that we observe.

If metallicity is the key physical variable driving the luminosity
trends, then the existence of radial metallicity gradients in galaxies
\citep[e.g.][]{1999PASP..111..919H} will complicate the analysis.
Studies generally show that the metallicity of galactic disks
decreases with increasing galactocentric radius, implying that a
simple segregation of SNe based on host galaxy stellar mass or
metallicity may (ultimately) require refinement to include, for
example, the location of the SN within the host. No trends are
apparent between SN residual and galactocentric radius in the SNLS
data. However, many factors complicate this analysis, the most
pernicious of which is that the measured SN positions must be
deprojected \citep[e.g.][]{2000ApJ...542..588I}.  This requires a
knowledge of the structural parameters of the galaxy (for example
inclination), difficult to measure for high-redshift galaxies using
ground-based data. Additionally, for SNe Ia with long delay times
between progenitor formation and explosion, the progenitor star may
migrate away from the region in which it is formed. Investigations of
this type must await higher-quality low-redshift data.

\section{Conclusions}
\label{sec:conclusions}

In this paper we have examined the photometric properties of Type Ia
Supernovae (SNe Ia) as a function of their host galaxy properties
using new data from the Supernova Legacy Survey (SNLS) and literature
data for SNe Ia at low-redshift and $z>1$. Our principal findings are

\begin{itemize}
\item As expected, SN Ia light-curve widths closely track the specific
  star-formation rate (sSFR) and stellar mass (\mstellar) of their
  host galaxies -- massive and/or low sSFR galaxies host SNe Ia with
  lower stretches (narrower light curves). There is only a mild
  dependence of SN Ia colour (\col) on the host galaxy -- SNe Ia in
  low sSFR galaxies are slightly bluer than those in high-sSFR host
  galaxies, but the colours of SNe Ia in high-\mstellar\ and
  low-\mstellar\ hosts are consistent.
\item SNe Ia in low sSFR host galaxies, and SNe Ia in massive host
  galaxies, are systematically brighter by 0.06-0.09\,mag ($>3\sigma$)
  \textit{after} light-curve shape and colour corrections. This is not
  dependent on any assumed cosmological model. Interpreting \mstellar\
  as an approximate metallicity indicator, this implies that
  metal-rich environments host brighter SNe Ia post-correction.
\item Galaxies with low sSFRs host SNe Ia with a smaller slope
  ($\beta$) between SN Ia luminosity and colour, and which have a
  smaller scatter on Hubble diagrams. This is dependent on the maximum
  colour of the SN which is considered -- the effect is more
  significant (2--$3\sigma$) when redder SNe are included. The slope
  of the relationship between stretch and luminosity is consistent for
  SNe Ia across different galaxy types.
\item The SN Ia luminosity variation with host properties introduces a
  systematic error into cosmological analyses, as the mean \mstellar\
  and sSFR of the hosts, and hence the mix of SNe Ia, evolves with
  redshift. This effect is exacerbated by the biased selection of
  existing low-redshift SNe Ia, and amounts to a systematic error on
  $w$ comparable to the statistical errors.
\item We propose that this be corrected for in all future SN Ia
  cosmological analyses by incorporating a host galaxy term into the
  fit. Specifically, we show that the use of two absolute magnitudes
  for SNe Ia, one for those in low-\mstellar\ (or low-metallicity)
  hosts, and one for events in more massive, metal-rich hosts, leads
  to an improvement in cosmological fits at 3.8--4.5$\sigma$, and
  removes the host dependence.
\item The SN Ia luminosity effects appear consistent with theoretical
  expectations of the dependence of SN Ia luminosities on progenitor
  metallicity. The effect of metallicity is predicted to translate
  into an offset in the effective stretch--luminosity relation, with
  brighter SNe in metal-rich environments after correction, as
  observed in our data.
\end{itemize}

These results are consistent with earlier observational studies which
have found evidence for SN Ia luminosity variation as a function of
host galaxy properties at 2--2.5$\sigma$ confidence
\citep{2008ApJ...685..752G,2009ApJ...700.1097H,2009arXiv0912.0929K}.
The SNLS dataset, spanning a larger range in host \mstellar\ and a
less biased selection of SNe Ia than at low redshift, provides tighter
constraints on host galaxy dependencies and broadly supports the
results of these earlier studies.

Redshift evolution in SN Ia properties driven by metallicity effects
in the SN Ia population has been suggested as a systematic error in SN
Ia cosmology for many years
\citep{1998ApJ...495..617H,2000ApJ...528..590H,2001ApJ...557..279D,2003ApJ...590L..83T,2009Natur.460..869K}.
The results of this paper detect this evolution, demonstrate that it
is qualitatively consistent with theoretical predictions, and show
that it can be corrected using supplementary information on the
environments in which the SNe explode. There are two key implications
from the perspective of measuring dark energy. The first is that
additional information on the SN Ia host galaxies, such as
multi-colour photometry covering a broad wavelength range, will be an
essential requirement for future SN Ia cosmological analyses and
surveys. Multi-colour, rolling searches similar to SNLS are obviously
well-placed to provide this information as a natural product of the
survey strategy.

The second implication is the urgent need for new low-redshift SN Ia
samples, where events are selected without regard to their host galaxy
type or brightness. Existing searches are mostly galaxy targeted --
repeatedly imaging the same catalogued galaxies -- and result in
heavily biased samples of SNe Ia. The Nearby Supernova Factory
\citep{2002SPIE.4836...61A}, and the next generation low-redshift
surveys such as the Palomar Transient Factory
\citep[PTF;][]{2009PASP..121.1334R,2009PASP..121.1395L} and SkyMapper
\citep{2007PASA...24....1K}, should provide samples of SNe Ia selected
in a similar way to those at higher redshift, reducing the systematic
uncertainties associated with evolving galaxy populations.

\section*{Acknowledgements}

MS acknowledges support from the Royal Society. This paper is based in
part on observations obtained with MegaPrime/MegaCam, a joint project
of CFHT and CEA/DAPNIA, at the Canada-France-Hawaii Telescope (CFHT)
which is operated by the National Research Council (NRC) of Canada,
the Institut National des Sciences de l'Univers of the Centre National
de la Recherche Scientifique (CNRS) of France, and the University of
Hawaii. This work is based in part on data products produced at the
Canadian Astronomy Data Centre as part of the CFHT Legacy Survey, a
collaborative project of NRC and CNRS.  Based in part on observations
obtained with WIRCam, a joint project of CFHT, Taiwan, Korea, Canada,
France, at the Canada-France-Hawaii Telescope (CFHT) which is operated
by the National Research Council (NRC) of Canada, the Institute
National des Sciences de l'Univers of the Centre National de la
Recherche Scientifique of France, and the University of Hawaii. This
work is based in part on data products produced at TERAPIX, the WIRDS
(WIRcam Deep Survey) consortium, and the Canadian Astronomy Data
Centre. This research was supported by a grant from the Agence
Nationale de la Recherche ANR-07-BLAN-0228.  Canadian collaboration
members acknowledge support from NSERC and CIAR; French collaboration
members from CNRS/IN2P3, CNRS/INSU and CEA.  Based in part on
observations made with ESO Telescopes at the Paranal Observatory under
program IDs 171.A-0486 and 176.A-0589.  Based in part on observations
obtained at the Gemini Observatory, which is operated by the
Association of Universities for Research in Astronomy, Inc., under a
cooperative agreement with the NSF on behalf of the Gemini
partnership: the National Science Foundation (United States), the
Science and Technology Facilities Council (United Kingdom), the
National Research Council (Canada), CONICYT (Chile), the Australian
Research Council (Australia), Ministério da Ciência e Tecnologia
(Brazil) and Ministerio de Ciencia, Tecnología e Innovación Productiva
(Argentina).  The programmes under which data were obtained at the
Gemini Observatory are: GS-2003B-Q-8, GN-2003B-Q-9, GS-2004A-Q-11,
GN-2004A-Q-19, GS-2004B-Q-31, GN-2004B-Q-16, GS-2005A-Q-11,
GN-2005A-11, GS-2005B-Q-6, GN-2005B-Q-7, GN-2006A-Q-7, and
GN-2006B-Q-10. Some of the data presented herein were obtained at the
W.M. Keck Observatory, which is operated as a scientific partnership
among the California Institute of Technology, the University of
California and the National Aeronautics and Space Administration. The
Observatory was made possible by the generous financial support of the
W.M. Keck Foundation. Based on observations made with the NASA/ESA
Hubble Space Telescope, obtained at the Space Telescope Science
Institute, which is operated by the Association of Universities for
Research in Astronomy, Inc., under NASA contract NAS 5-26555.

\bibliographystyle{mn2e}

\label{lastpage}

\end{document}